%% file: template.tex
\documentclass{article}

\usepackage{arxiv}

\usepackage[utf8]{inputenc} % allow utf-8 input
\usepackage[T1]{fontenc}    % use 8-bit T1 fonts
\usepackage{hyperref}       % hyperlinks
\usepackage{url}            % simple URL typesetting
\usepackage{booktabs}       % professional-quality tables
\usepackage{amsfonts}       % blackboard math symbols
\usepackage{nicefrac}       % compact symbols for 1/2, etc.
\usepackage{microtype}      % microtypography
\usepackage{lipsum}
\usepackage{graphicx}
\graphicspath{ {./images/} }

\usepackage{cite}
\usepackage{amsmath,amssymb,amsfonts}
\usepackage{array}
\usepackage{algorithmic}
\usepackage{graphicx}
\usepackage{textcomp}
\usepackage{xcolor}
\usepackage{enumitem}
\usepackage{tabularx}
\usepackage{threeparttable}
\usepackage{adjustbox}
\usepackage{threeparttable}
\usepackage{siunitx}
\usepackage{url}
\usepackage{pifont}     % for checkmarks and crosses
\def\BibTeX{{\rm B\kern-.05em{\sc i\kern-.025em b}\kern-.08em
    T\kern-.1667em\lower.7ex\hbox{E}\kern-.125emX}}
\usepackage{tikz}
\usepackage{makecell}
\usepackage{subcaption}
\usepackage{booktabs}
\usepackage{tablefootnote}

\newcommand{\sysname}{\textit{RACAM}}
\newcommand*\circled[1]{\tikz[baseline=(char.base)]{
            \node[shape=circle,fill,inner sep=2pt] (char) {\textcolor{white}{#1}};}}

\title{\textbf{\textit{\sysname}}: Enhancing DRAM with Reuse-Aware Computation and Automated Mapping for ML Inference}

\author{
 Siyuan Ma \\
  Department of Electrical and Computer Engineering\\
  University of Texas at Austin\\
  Austin, TX 78712 \\
  \texttt{siyuan.ma@utexas.edu} \\
   \And
 Jiajun Hu \\
  School of Computing and Augmented Intelligence\\
  Arizona State University\\
  Tempe, AZ 85281  \\
  \texttt{jiajunh5@asu.edu} \\
  \And
 Jeeho Ryoo \\
  Olsen College of Engineering and Science\\
  Fairleigh Dickinson University\\
  Vancouver\\
  \texttt{j.ryoo@fdu.edu} \\
  \And
 Aman Arora \\
  School of Computing and Augmented Intelligence\\
  Arizona State University\\
  Tempe, AZ 85281\\
  \texttt{aman.kbm@asu.edu} \\
  \And
 Lizy Kurian John \\
  Department of Electrical and Computer Engineering\\
  University of Texas at Austin\\
  Austin, TX 78712 \\
  \texttt{ljohn@ece.utexas.edu} \\
}

\begin{document}
\maketitle
\input{sec/00_abstract}

\input{sec/01_intro}

\input{sec/02_background}
\input{sec/03_architecture}

\input{sec/04_mapping}

\input{sec/05_methodology}

\input{sec/06_results}
\input{sec/07_discussion}
\input{sec/08_related_work}

\input{sec/09_conclusion}

%%%%%%% -- PAPER CONTENT ENDS -- %%%%%%%%

%%%%%%%%% -- BIB STYLE AND FILE -- %%%%%%%%
\newpage
\bibliographystyle{IEEEtranS}
\bibliography{references}

\end{document}

%% file: sec/00_abstract.tex
\begin{abstract}
% Processing-In-Memory (PIM) architectures have emerged as a promising solution for accelerating machine learning, particularly due to the rapidly increasing computational demands and data transfer overhead between the host and device in modern large language models (LLMs). However, most prior PIM works primarily focus on the decoding stage of LLMs and lack support for end-to-end evaluation. In addition, their workload mapping schemes are highly constrained, with limited exposure to the data reuse of general matrix-matrix multiplication (GEMM). These shortcomings make most PIM simulators and architectures neither comprehensive nor robust enough to support diverse workload mapping strategies. To this end, we propose \textbf{\textit{\sysname}}, a reuse-aware in-DRAM bit-serial architecture that achieves performance comparable to GPUs while providing robust architectural and framework support for flexible workload mapping schemes. By using \textbf{\textit{\sysname}}, the optimal combination of architecture parameters and workload mapping strategies is achieved in a co-design approach. We show that for a given LLM, \textbf{\textit{\sysname}} delivers up to 150x speedup against H100 with less than 2\% DDR5 area overhead.  
In-DRAM Processing-In-Memory (DRAM-PIM) has emerged as a promising approach to accelerate memory-intensive workloads by mitigating data transfer overhead between DRAM and the host processor. Bit-serial DRAM-PIM architectures, further enhance efficiency by supporting runtime variable data precision, which is critical for emerging workloads, such as large language model (LLM) inference. However, existing works still have major limitations: 
%\\JR{abstract shoulnd't contain such extensive detailed listing. Just mention 4 things. we explain those later}
%(i) 
lack of data reuse, 
% Each bit stored in a DRAM row must be activated multiple times during multiplication, leading to $O(n^2)$ activations comparing to $O(n)$ with full bit-reuse.
% and each operand needs to be accessed multiple times during matrix multiplication, resulting in long latency.  
%(ii) 
significant amounts of redundant data transfer,  
% Fully utilizing DRAM's massive parallelism requires data duplication at several DRAM hierarchical levels. However, prior work relies on explicit data placement by the host CPU to duplicate data, resulting in redundant data movement with latency costs.
% (iii) compromised DRAM Reliability, and
% To achieve flexible precision, prior work rely on DRAM's charge-sharing mechanism to achieve in-DRAM computation. However, such an approach leads to unstable bitline voltage and current, is sensitive to DRAM vendors, and leads to reliability issues.
%(iii) 
and insufficient support for workload mapping.
To address these issues, we propose \sysname, the first in-DRAM bit-serial architecture which uses dedicated locality buffers, bit-serial PEs, popcount reduction units and broadcast units to enable data reuse and alleviate redundant data transfers.
Furthermore, a workload mapping mechanism is proposed to fully explore the massive parallelism of DRAM architecture and identify the best mapping scheme of a given workload.
We evaluate \sysname ~against GPUs and the state-of-the-art, in-DRAM PIM system, Proteus, across end-to-end LLM inferences.
% With only $6\%$ DDR5 memory chip area overhead, 
\sysname ~achieves $9\times$ to $102\times$ speedup over GPUs and $233\times$ higher performance per mm\textsuperscript{2} compared to Proteus in case of GPT3.
% Comparing to GPU, \sysname with 1TB DRAM achieves $16.88\times$ average speedup. Comparing to Proteus, \sysname achieves $205.73\times$ average performance per $mm^2$. \sysname incurs $6\%$ area overhead on DRAM chips.
\end{abstract}

%% file: sec/01_intro.tex
\section{Introduction}
% \textcolor{red}{\begin{itemize}
%     \item Mention DRAM industry trend of DRAM-PIM.
%     \item LLMs and AI are getting memory-bound and consist of GEMM/GEMV.
%     \item Existing DRAM-PIMs cannot do GEMM or it is very slow.
%     \item Mention concrete numbers of existing DRAM-PIM drawbacks (slow against GPU ..) and explain the reasons/solutions we provided.
% \end{itemize}}

As AI models continue to grow in scale and complexity, the demand for sustained data movement between memory and compute units has increased substantially. In large language models (LLMs), whose dominant operators are general matrix–matrix and matrix–vector multiplications (GEMM/GEMV), performance is fundamentally constrained by memory bandwidth rather than by the raw compute capability of GPU compute cores. Even with large GPU clusters equipped with high-bandwidth memory, LLM throughput fails to scale proportionally because the memory subsystem cannot supply data at the rate required by these operators, making memory the primary performance and energy bottleneck in contemporary LLM workloads \cite{flashAttention, efficient_scaling_transformer_inference, deepseepd}. In response, major memory vendors including Samsung, SK Hynix, and Micron are developing processing-in-memory (PIM) and processing-near-memory (PNM) architectures to more efficiently accelerate memory-bound workloads.\cite{samsung_pim,hynix_pim,newtown_AiM, micron_pim}.

% As AI continues to evolve rapidly, the demand for computing power is growing at an unprecedented rate. As a core component of modern AI applications, large language models (LLMs) urgently require increasing amounts of computing resources. Even with giant GPU clusters, coupled with high-bandwidth memory, LLMs' performance does not scale with the increasing computing power. As the core operators of LLMs are general matrix-matrix/vector multiplications (GEMM/GEMV), the bandwidth between memory and computing cores limits the system performance\cite{deepseepd,efficient_scaling_transformer_inference,flashAttention}, known as the memory bottleneck. Moving computing units closer to the memory, while utilizing significantly higher in-memory and near-memory bandwidth, is considered an effective architectural solution to alleviate the memory bottleneck. Modern industry memory vendors like Samsung, SK Hynix, and Micron are developing processing-in-memory (PIM) and processing-near-memory (PNM) architectures to accelerate memory-bound workloads\cite{samsung_pim,hynix_pim,newtown_AiM, micron_pim}. 

Beyond limitations imposed by memory bandwidth, precision flexibility has become an increasingly important requirement for modern workloads such as DNNs and LLMs\cite{yu2021anyprecisiondeepneuralnetworks, flexpe, 10.1145/3729215, slimllm}, which exhibit varying numerical sensitivity across layers and therefore benefit from architectures that adapt numerical precision to each layer’s error tolerance and algorithmic requirements. To support such variability, prior work has explored several precision-scalable PIM mechanisms including bit-serial, bit-sliced, and mixed-precision designs. Among these options, bit-serial PIM is particularly appealing due to its natural scaling with DRAM row parallelism and ability to adjust effective precision through serialization. Recent Processing-Using-DRAM (PUD) systems\cite{computeDRAM,simdram, mimdram, proteus} demonstrate that DRAM arrays can execute massively parallel bit-serial logic operations by exploiting charge sharing and sense-amplifier behavior within largely standard DRAM ACT/PRE command sequences, enabling high area efficiency and improved performance-per-mm\textsuperscript{2} for memory-bound workloads.

\begin{figure}
\centering
    \includegraphics[width= 0.5\columnwidth]{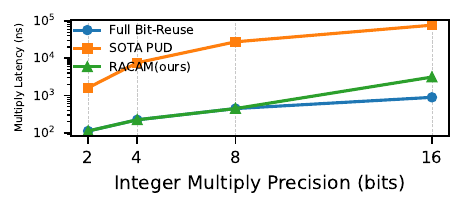}
    % \vspace{-0.3in}
    \caption{Integer Multiplication Latency}
    \label{mul_latency}
\end{figure}

Existing PUD systems, however, exhibit long multiplication latencies because all multiplicand bits must be accessed for every bit of the multiplier, resulting in O(n²) quadratic scaling latencies for an n-bit integer multiply\cite{proteus}. In DRAM-based bit-serial logic, each operand-bit access typically incurs a row activation and precharge, so repeatedly revisiting the same operand bits generates a large number of ACT–PRE cycles. This inefficiency arises from the lack of bit-level locality in current PUD data mappings as operand bits are not positioned to allow reuse, forcing the system to re-read the same bits for every partial product. Fig.~\ref{mul_latency} quantifies this effect. The orange curve (“SOTA PUD”) shows a design with no bit reuse, where multiplication latency grows steeply because every multiplicand bit is fetched independently for each multiplier bit. The blue curve represents an idealized design with full bit reuse, where operand bits are accessed once and reused across all partial products, yielding far lower latency growth.  Overall, the figure highlights that excessive row activations caused by bit-level non-locality are the dominant contributor to multiplication latency in existing PUD systems.

Another challenge in PUD systems is the cost of transferring dynamic data across the DRAM hierarchy. While DRAM PIMs and PUDs reduce data movement to  central processing units, there is need for significant data transfer in intermediate steps. While static operands (such as weights) can be pre-duplicated offline in whatever pattern that maximizes parallelism, dynamic operands such as activations, intermediate data, or kernel inputs must be replicated at runtime to columns or bank that participates in parallel computation. For example, if an input matrix tile $A$ must be available in all banks to enable parallel partial-product generation, prior PUD systems such as ComputeDRAM and SIMDRAM rely on the host processor to explicitly write $A$ into each bank, incurring roughly $\#\text{Banks} \times \text{Bytes}_A$ of data movement across the memory channel~\cite{computeDRAM,simdram}. Although DRAM provides high internal bandwidth, this replication is performed over the off-chip CPU--DRAM interface, so the overhead is again dominated by limited channel bandwidth and increased energy per transferred byte. In contrast, providing an internal DRAM-supported broadcast pathway would reduce the required off-chip transfer to only $\text{Bytes}_A$, allowing the duplication to occur within DRAM’s high-bandwidth internal fabric and significantly lowering data-movement overhead.

In addition, the design of systematic workload–to–DRAM mappings remains insufficiently generalized in prior DRAM-based PIM research. Existing works do incorporate mapping strategies, but they are typically tightly coupled to specific operators, dataflows, or DRAM organizations such as hand-crafted layouts for bit-parallel logic in SIMDRAM and Proteus~\cite{simdram,proteus}, fixed mapping patterns designed around a particular computation model~\cite{newtown_AiM}, or heuristic, architecture-specific placement policies as in MIMDRAM~\cite{mimdram}. While these approaches enable their respective mechanisms, they do not provide a general mapping framework that exposes the full design space. We envision that the GEMM mapping spaces can be formalized as dimensions mapped to DRAM hierarchical structures, and analytical performance models can be used to automatically search through the entire space and generate the mapping. The mappings explored in prior work tend to be largely manual, narrow in scope, limiting portability across different DRAM configurations and reducing adaptability to diverse or dynamically changing workload dimensions.

% Furthermore, workload mapping remains an underexplored area in prior DRAM PIM works\JR{I think prior work did some mapping to enable bit parallel. I think this statement is flawed}. Existing studies either rely on hand-crafted, workload- and architecture-specific mappings \cite{simdram, proteus}, define fixed mapping schemes \cite{newtown_AiM}, or propose heuristic approaches tailored to specific architectures \cite{mimdram}. None provides a formalized mapping space that decouples mapping from workload and architectural details\JR{what is so special about formalizing it? We decouple mapping? We coupled it with memory configuration, no?}. Consequently, the explored mappings are limited, constraining adaptability to diverse and dynamic workload shapes and architectures.

To address these limitations, we propose \sysname, a scalable in-DRAM bit-serial PIM architecture that enables high-throughput and reliable computation for AI workloads by jointly improving bit-level reuse, operand locality, data broadcasting, and workload mapping. \sysname ~introduces bit-serial processing elements with locality buffers that significantly reduce redundant ACT--PRE operations by reusing operand bits across partial products. This leads to \sysname~ approaching the ideal behavior in Fig. \ref{mul_latency} (shown in green).
To minimize host-to-DRAM traffic, \sysname ~integrates an internal broadcast unit that replicates dynamic operands across banks entirely within DRAM's high-bandwidth fabric, avoiding costly off-chip transfers. In addition, \sysname ~provides a generalized mapping framework that separates mapping decisions from both workload structure and DRAM organization, enabling systematic exploration of the mapping space across diverse DRAM configurations. Together, these architectural components allow \sysname ~to efficiently accelerate end-to-end AI workloads while respecting DRAM reliability constraints and preserving area efficiency.

Specifically, this work makes the following contributions:
\begin{itemize}
\item We present \sysname, an in-DRAM bit-serial PIM architecture that integrates bit-serial PEs with locality buffers for scalable AI acceleration.
\item We reduce redundant row activations by enabling bit-level operand reuse across partial products.
\item We introduce an internal DRAM broadcast unit that replicates dynamic operands across banks without host involvement, reducing data transfer overheads.
\item We provide a generalized mapping framework that decouples workload structure from DRAM configuration to explore a broad mapping space.
% \item We demonstrate up to $112\times$ speedup over H100 and $205.73\times$ higher performance per~mm\textsuperscript{2} than Proteus with only $6\%$ area overhead.
\end{itemize}

\noindent\sysname ~delivers an average of 90.1x and 15.6x performance improvement over GPUs on two end-to-end inference senarios with GPT3 and Llama3. Likewise, \sysname~ achieves an average of $46\times$ over GPUs for performance/$mm^2$ and  improvement (geomean) of more than $200\times$  over SOTA PIMs with only approximately 4\% chip area overhead. The end-to-end throughput improvement over GPUs is significant for decode operations in transformers, but not for prefill operations. Ablation studies showed that, among the added features, locality buffer yields the biggest improvement.

The remainder of this paper is organized as follows.
Section~\ref{sec:background} introduces the necessary background on DRAM organization, bit-serial computation, and matrix-multiplication tiling and mapping.
Section~\ref{sec:arch} describes the proposed \sysname ~architecture and microarchitectures.
Section~\ref{sec:mapping} presents \sysname’s mapping and scheduling framework.
Sections~\ref{sec:methodology} and~\ref{sec:results} detail our evaluation methodology and experimental results.
Finally, Sections~\ref{sec:discussion}, \ref{sec:related_work}, and \ref{sec:conclusion} provide discussion, related work, and concluding remarks.

%% file: sec/02_background.tex
\section{Background}
\label{sec:background}
\subsection{DRAM Organization}

% Fig.\ref{dram_overview} shows an overview of the DRAM system. The DRAM hierarchy starts at \textit{channels}, \textit{ranks}, \textit{devices}, \textit{banks}, and extends to the lowest-level \textit{arrays}.\cite{memory_system_book, ambit, simdram, mimdram} Each channel is divided into multiple ranks, and each rank is divided into multiple devices. Each device contains multiple banks, and each bank contains multiple arrays. The bit cells inside an array are accessed by the address decoder, and the 1-bit data is encoded as the voltage level of the storage capacitor inside the bit cell. To read the data, the \textit{wordline} is first asserted and all cells along the wordline are connected to the corresponding \textit{bitlines}. A voltage difference between the reference voltage and cell capacitor is then  sensed by the \textit{sense amplifiers}  and amplified to a digital value to be stored inside the row buffer. To write back, the wordline still needs to be asserted, and the connected bitlines will be charged or discharged based on the value stored inside the sense amplifiers.  

\begin{figure}
\centering
    \includegraphics[width= 0.6\columnwidth]{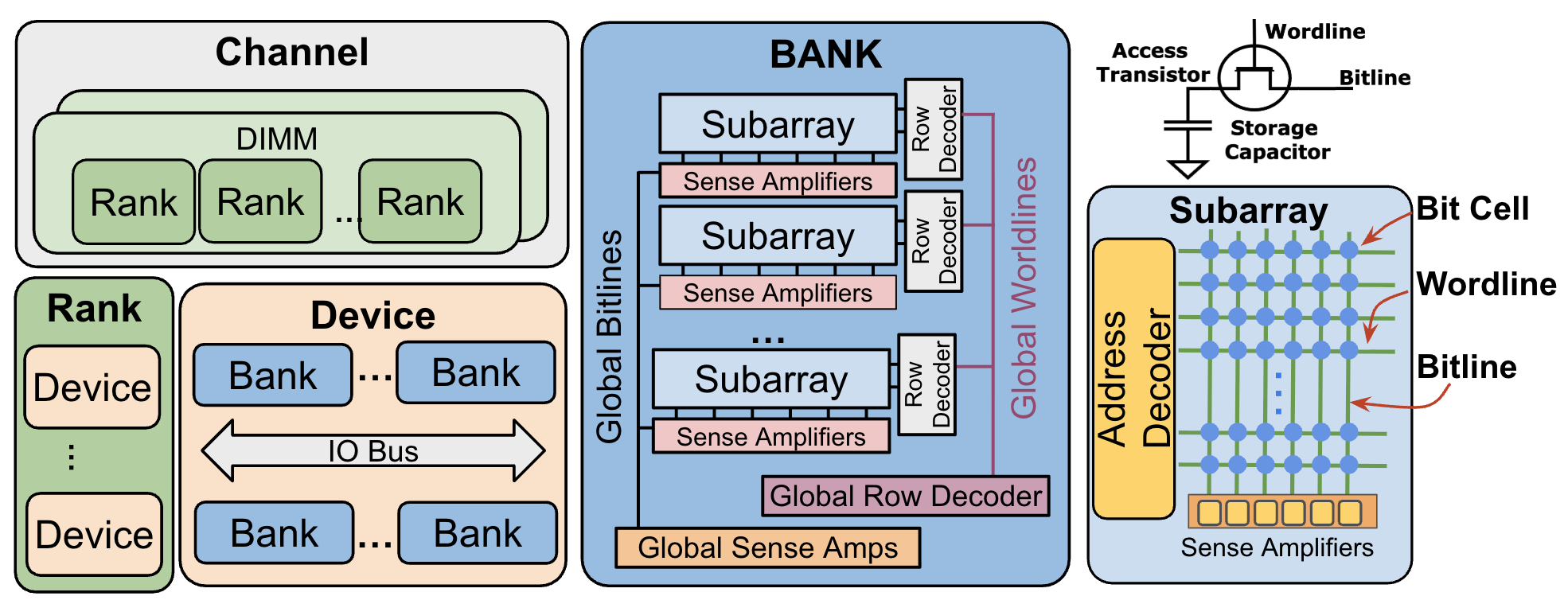}
    \caption{DRAM organization overview.}
    \label{dram_overview}
\end{figure}

Fig.~\ref{dram_overview} illustrates the hierarchical organization of a modern DRAM system. The hierarchy spans from \textit{channels} to \textit{ranks}, \textit{devices}, \textit{banks}, and ultimately \textit{sub-arrays}.\cite{memory_system_book, ambit, simdram, mimdram} Each channel contains multiple ranks; each rank comprises multiple devices; each device is divided into banks; and each bank consists of multiple sub-arrays.

Within an array, data is stored in individual bit cells, where each bit is represented by the charge level of a storage capacitor. Accessing a row begins by asserting the \textit{wordline}, which connects all cells in that row to their corresponding \textit{bitlines}. The small voltage perturbation contributed by each cell is sensed and amplified by the \textit{sense amplifiers}, which restore the full digital value and place the resulting data into the row buffer. Writing data back similarly requires asserting the wordline; the sense amplifiers then drive the bitlines to charge or discharge the capacitors according to the stored value.
\subsection{Bit-serial Computation}
% In contrast to bit-parallel scheme, bit-serial scheme compute one bit of operand per cycle rather than the full-precision operand. Generally, bit-serial schemes perform in a SIMD fashion and can achieve even higher throughput than bit-parallel scheme when operating on high-precision data width. However, bit-serial schemes requires a transposed vertical data layout per DRAM subarray column, which can be achieved by a transpose unit. \cite{simdram, mimdram, proteus, pimsab} Static data like weights can be transposed and written to DRAM prior to runtime execution as a one-time effort. \cite{ambit, simdram, mimdram, proteus} The area overhead of attached PEs are discussed in section \ref{sec:area_estimation}. We refer to \cite{mimdram,simdram,comefa} for further explanation on bit-serial computation.

In contrast to bit-parallel execution, a bit-serial scheme processes one bit of each operand per cycle rather than the full-precision value. Bit-serial computation is typically implemented in a SIMD manner and can offer higher effective throughput than bit-parallel designs, and supports runtime variable data precisions. To support bit-serial operations in DRAM, data must be stored in a vertically transposed layout, aligning bits of the same significance across DRAM subarray columns. This layout is usually generated by a transpose unit.\cite{simdram, mimdram, proteus, pimsab}

Static data such as model weights can be pre-transposed and written to DRAM offline as a one-time preprocessing step.\cite{ambit, simdram, mimdram, proteus} 
% The area overhead of the attached processing elements (PEs) is evaluated in Section~\ref{sec:area_estimation}. 
For a more detailed discussion of bit-serial arithmetic and its use in DRAM-PIM architectures, we refer readers to prior works\cite{mimdram, simdram, comefa}.

% \begin{figure}
%     \includegraphics[width= \columnwidth]{figs/bit_serial_op_Layer 1.pdf}
%     \caption{Example of two 4-bit integer operands in-DRAM bit-serial operation, reproduced from \cite{pimsab}.}

%     \label{bit_serial}
% \end{figure}

% Fig.~\ref{bit_serial} shows a basic example of 4-bit integer operation using bit-serial arithmetic. In bit-serial computation, data is stored in a transposed format, where the operands A, B, and C are organized vertically within memory cells. In this example, operands A and B are stored from $row_i$ to $row_{i+3}$ and $row_j$ to $row_{j+3}$, respectively, where $A[0]$ to $A[3]$ and $B[0]$ to $B[3]$ represent the four bits of each operand. Accordingly, $C[0]$ to $C[3]$ represent the 4 bits of the processed result; the rest bits are not shown here. During the memory read operation, two wordlines containing a bit of each operand are activated. The processing element performs the 1-bit computation and writes the resulting bit back to a designated DRAM wordline. In here, $A[0]$ and $B[0]$ are first loaded into the row buffer, and the processed bit is written to $row_k$ as $C[0]$. If is addition, carry bits are propagated through the processing elements and eventually contribute to the MSB.

% \textcolor{blue}{
\subsection{Tiling \& Mapping of Matrix Multiplication}
\begin{figure}
\centering
    \begin{subfigure}{0.3\columnwidth}
        \includegraphics[width= \columnwidth]{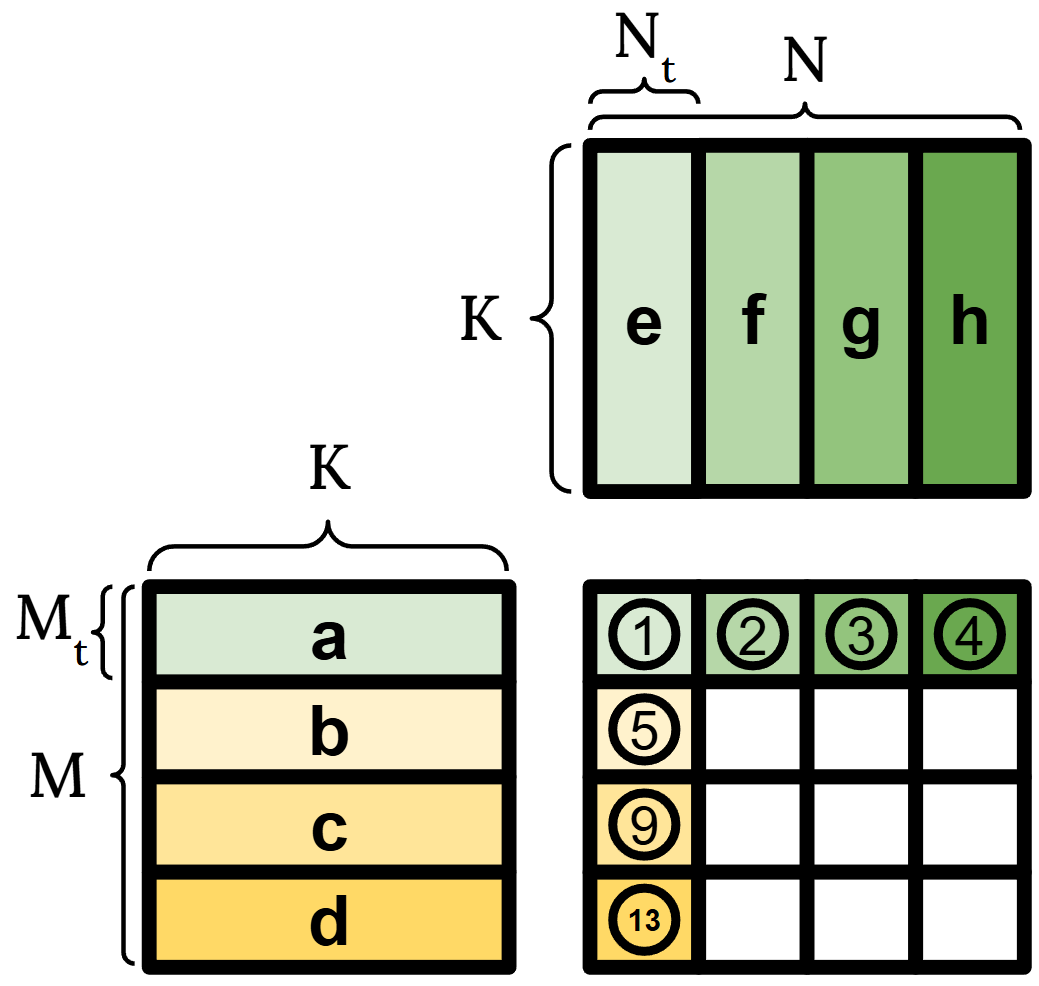}
        \caption{Tiling on M, N dimensions}
        \label{tiling-mn}
    \end{subfigure}
    \begin{subfigure}{0.3\columnwidth}
        \includegraphics[width= \columnwidth]{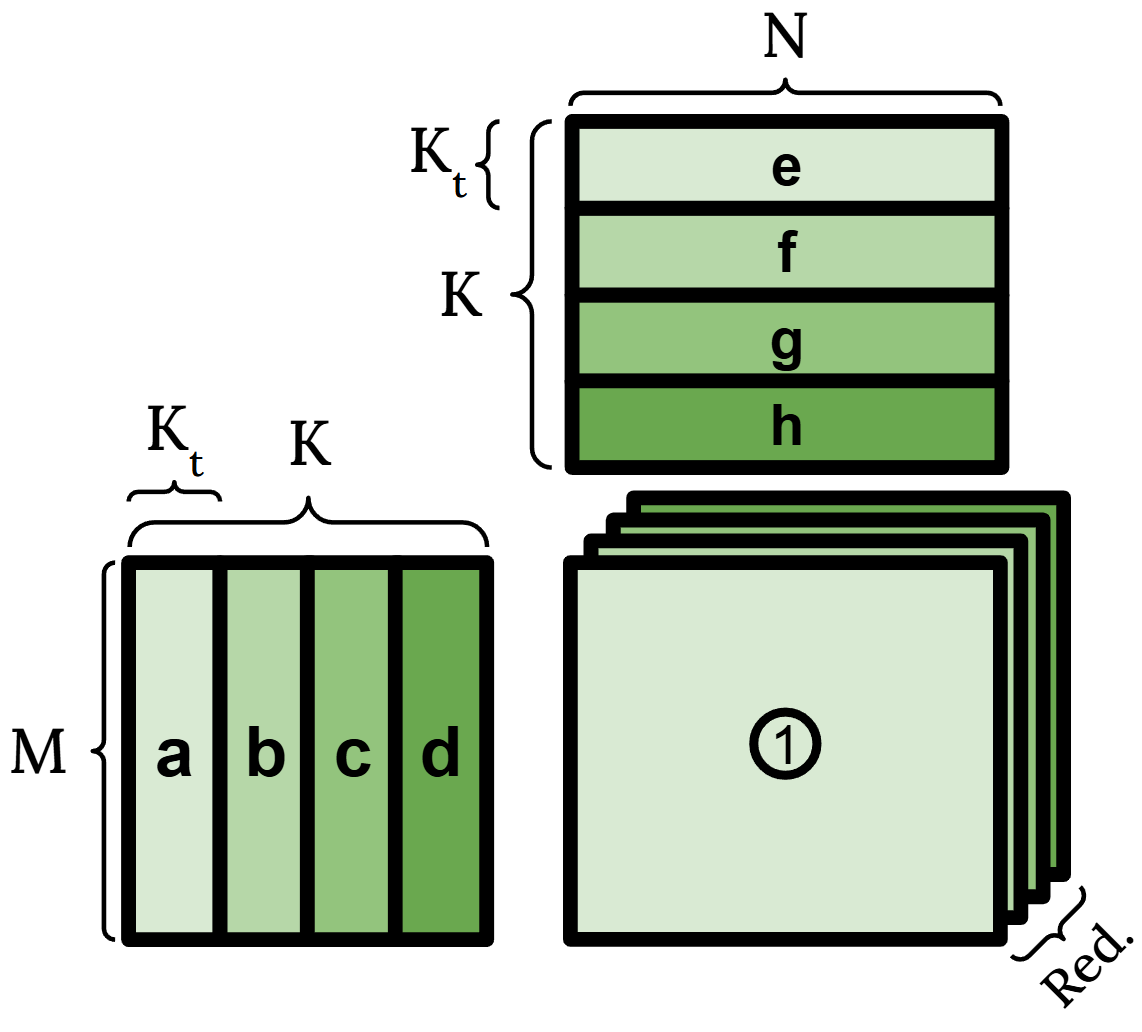}
        \caption{Tiling on K dimension, requiring reduction}
        \label{tiling-k}
    \end{subfigure}
    \caption{Tiling of Matrix Multiplication}
    \label{matmul-tiling}
\end{figure}

Fig.~\ref{matmul-tiling} illustrates representative tiling strategies for matrix multiplication, where tiling choices dictate the amount of data movement, the degree of data duplication, and the level of parallelism exploitable by the system. Each processor computes one tile (shown in different colors) in parallel with others.

Tiling along the $M$ or $N$ dimensions introduces \texttt{data duplication}. In Fig.~\ref{tiling-mn}, the $M$ and $N$ dimensions are partitioned into tiles of size $M_t$ and $N_t$, respectively. Computing the four output tiles (1–4) requires the input tiles $(a,e)$, $(a,f)$, $(a,g)$, and $(a,h)$, implying that tile $a$ must be replicated across four processors when each output tile is assigned to a single processor. Similarly, tile $e$ is duplicated across processors \#1, \#5, \#9, and \#13.

Tiling along the $K$ dimension, in contrast, creates partial outputs and introduces \texttt{reduction}. In Fig.~\ref{tiling-k}, the $K$ dimension is partitioned into $K_t$, and each processor produces a partial sum of one output tile. A reduction across the four processors is then required to accumulate these partial results into the final output matrix. Tiling across $M$, $N$, and $K$ can be combined to explore a larger space of parallelization and data-movement trade-offs.

In the context of in-DRAM processing, DRAM’s hierarchical organization naturally corresponds to multiple levels of parallel processors. For example, four ranks can be viewed as coarse-grained processors, while the sixteen banks within each rank behave as fine-grained parallel sub-processors. Mapping matrix tiles to this hierarchy is more involved, as each hierarchy level can be paired with a distinct tiling dimension. The sensitivity of tiling choices and hierarchy-aware mappings is further examined in Section~\ref{sec:mapping}.

%% file: sec/03_architecture.tex
\section{\sysname ~Architecture}
\label{sec:arch}
\input{tables/pim_cmd}
\begin{figure}
\centering
    \includegraphics[width= 0.6\columnwidth]{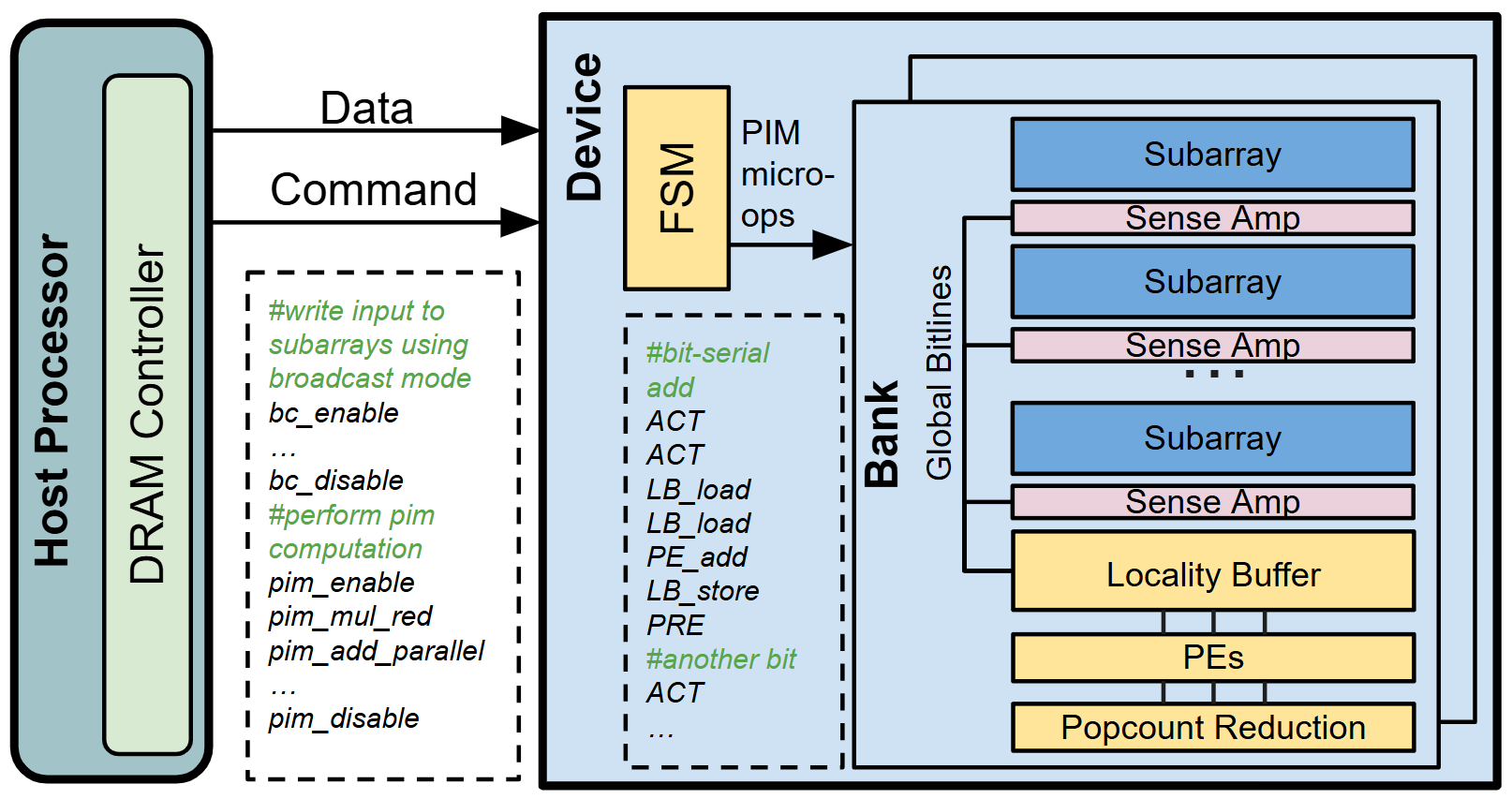}
    % \vspace{-0.2in}
    \caption{\sysname~system overview. Added peripherals are colored in yellow; broadcasting units are not shown.}
    \label{sys_overview}
\end{figure}

\begin{figure*}
     \centering
     \begin{subfigure}[t]{0.3\textwidth}
        \centering
        \includegraphics[width= \columnwidth]{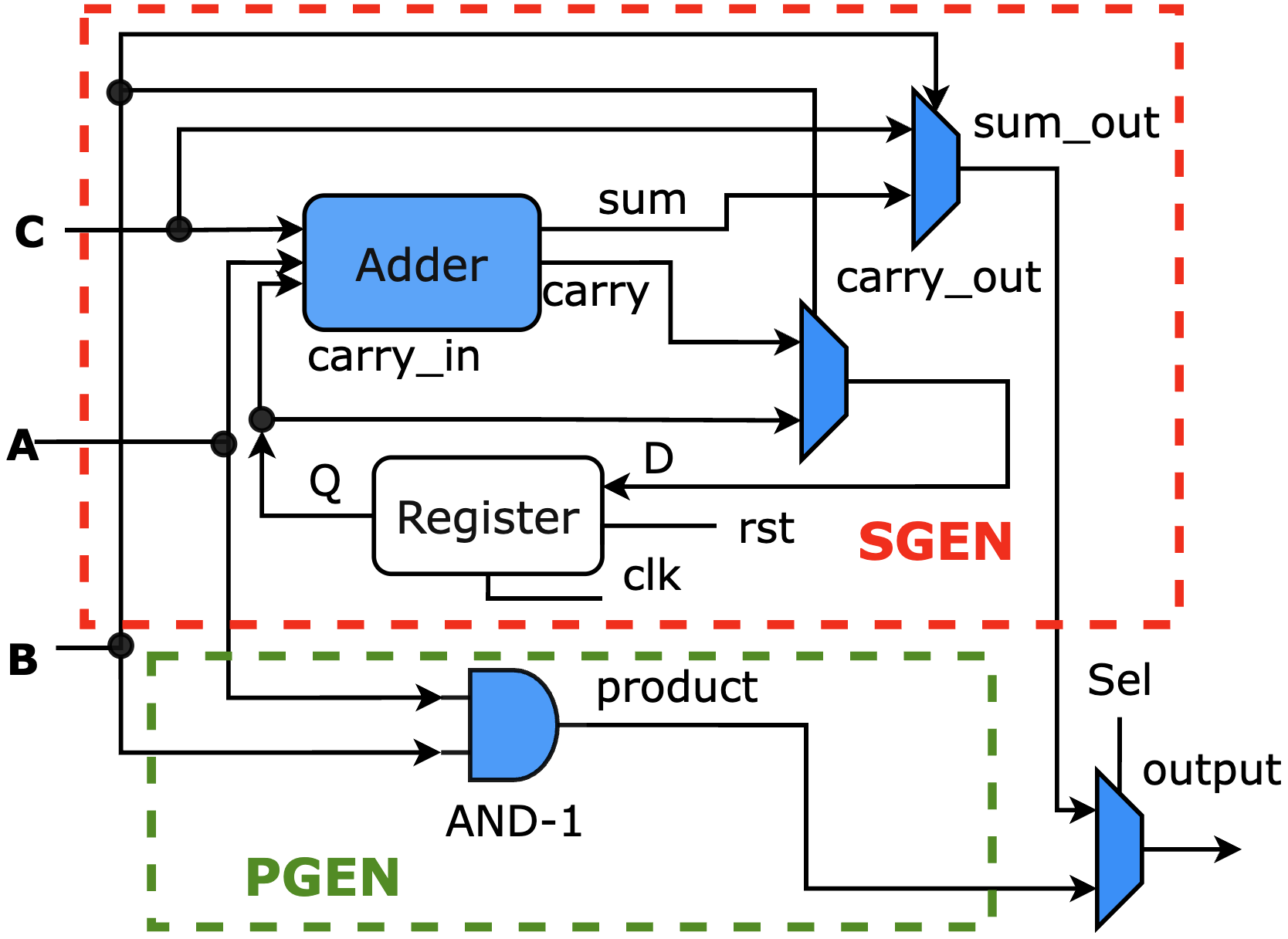}
        \caption{Bit-serial PE attached to each column of DRAM subarray. \textit{Product} is generated by \textbf{\textit{PGEN}}, and \textit{Sum} is generated by \textbf{\textit{SGEN}}.}
        \label{bit_serial_pe}
     \end{subfigure}
     % \begin{subfigure}{0.30\textwidth}
     \begin{subfigure}[t]{0.3\textwidth}
        \centering
        \includegraphics[width=0.7\columnwidth]
        {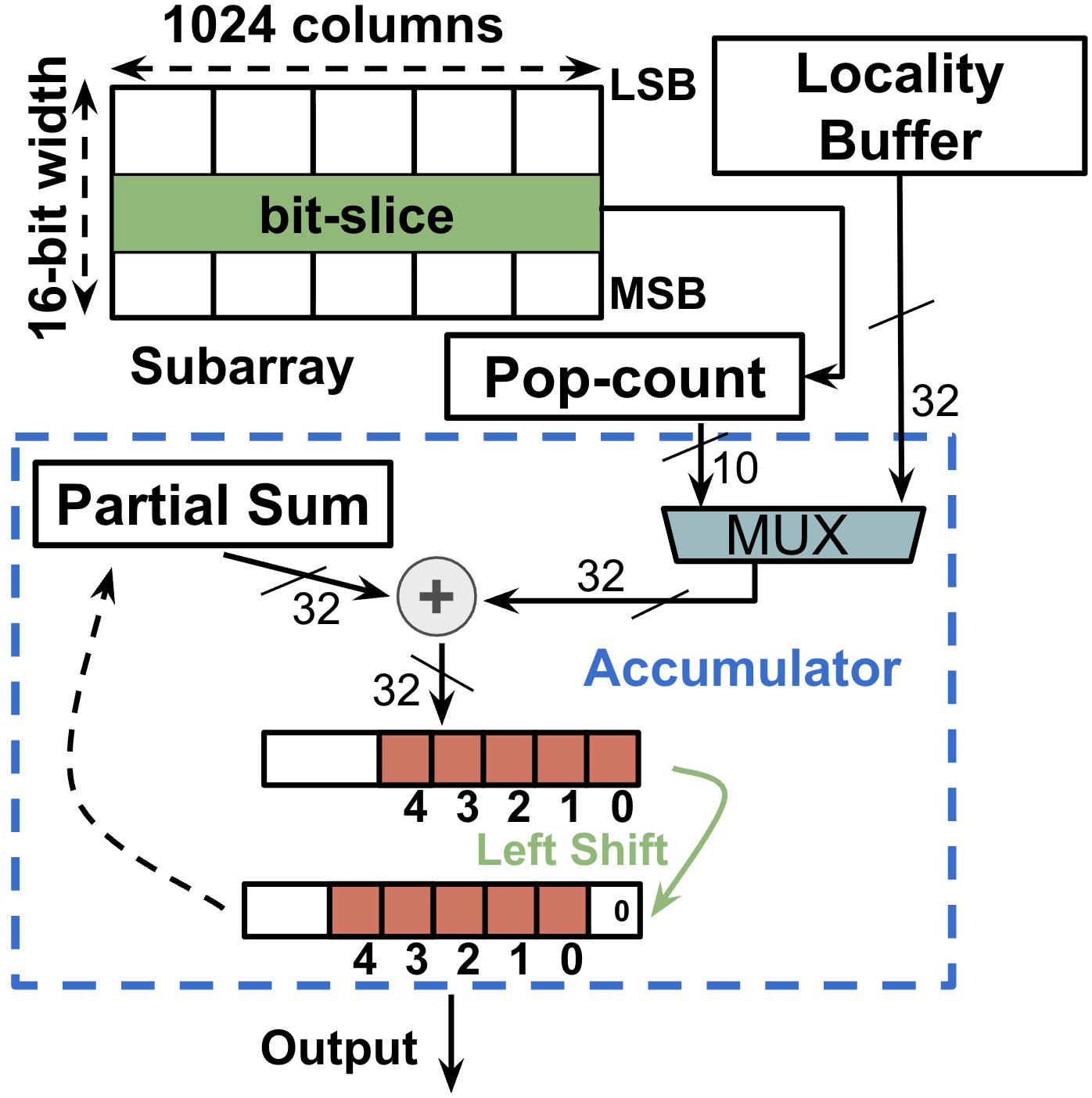}
        \caption{Popcount Reduction unit for 16-bit partial sum with 1024 subarrays.}
        \label{popcount_adder}
     \end{subfigure}
     \begin{subfigure}[t]{0.3\textwidth}
        \centering
        \includegraphics[width=\columnwidth]{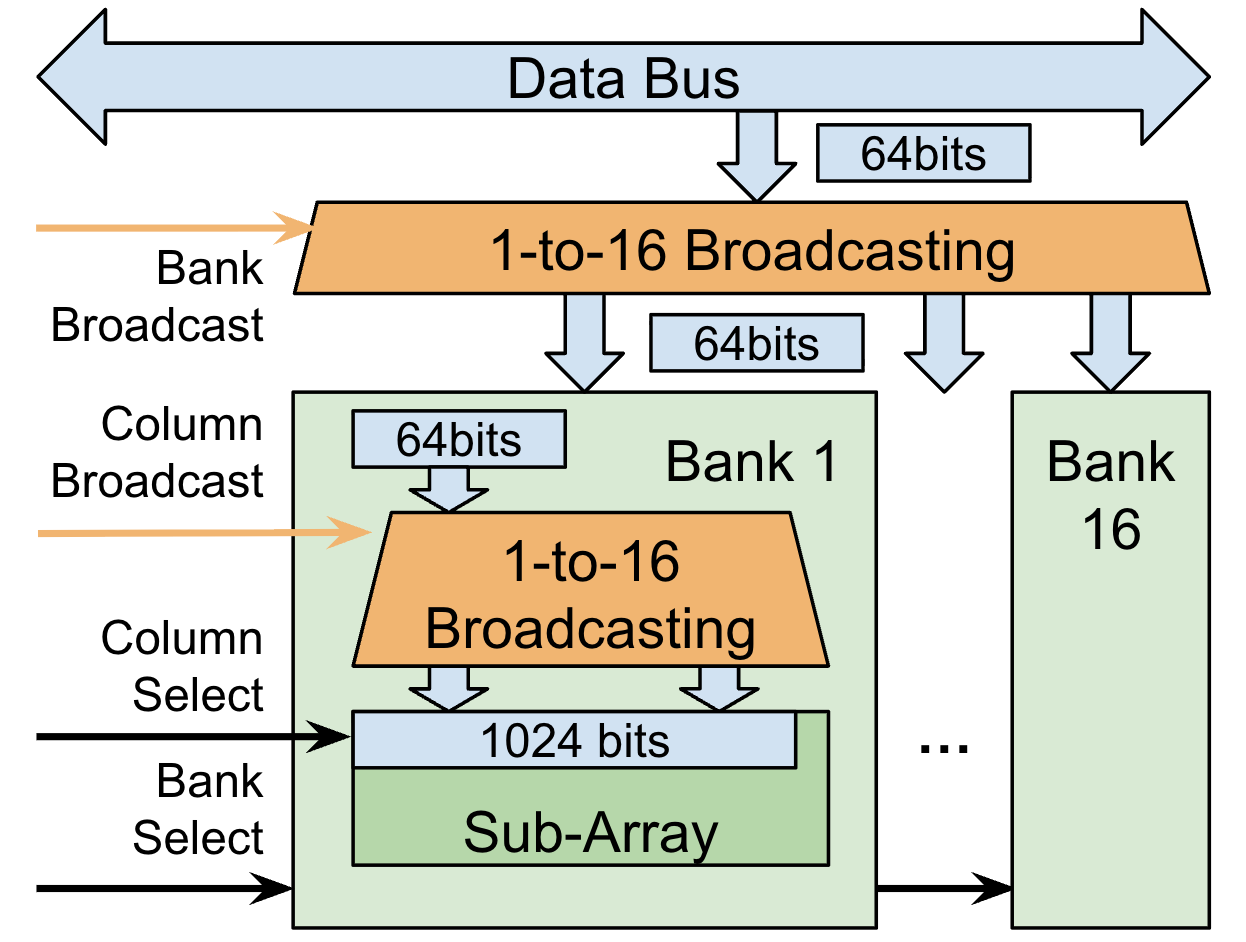}
        \caption{Broadcasting network for configuration of x8 devices, 16 banks and 64 subarrays.}
        \label{broadcast_unit}
     \end{subfigure}
     \label{fig:microarch}
    \caption{Added peripheral units to \sysname architecture}
\end{figure*}

\begin{figure}
     \centering
     \begin{subfigure}[t]{0.45\columnwidth}
     % \begin{subfigure}{0.35\textwidth}
        \centering
         \includegraphics[width= \columnwidth]{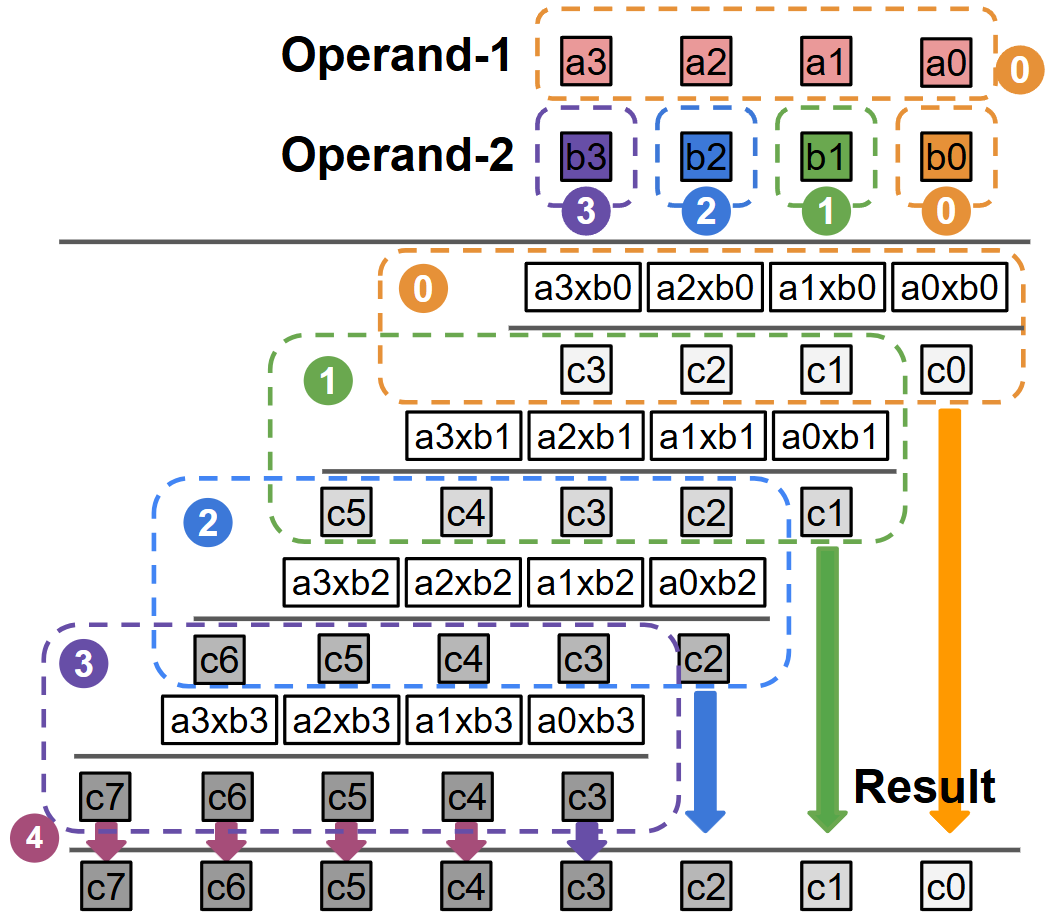}
        \caption{Multiplication Logistics and Steps}
        \label{fig:bit_serial_compute}
     \end{subfigure}
     % \begin{subfigure}{0.30\textwidth}
     \begin{subfigure}[t]{0.35\columnwidth}
        \centering
        \includegraphics[width= \columnwidth]{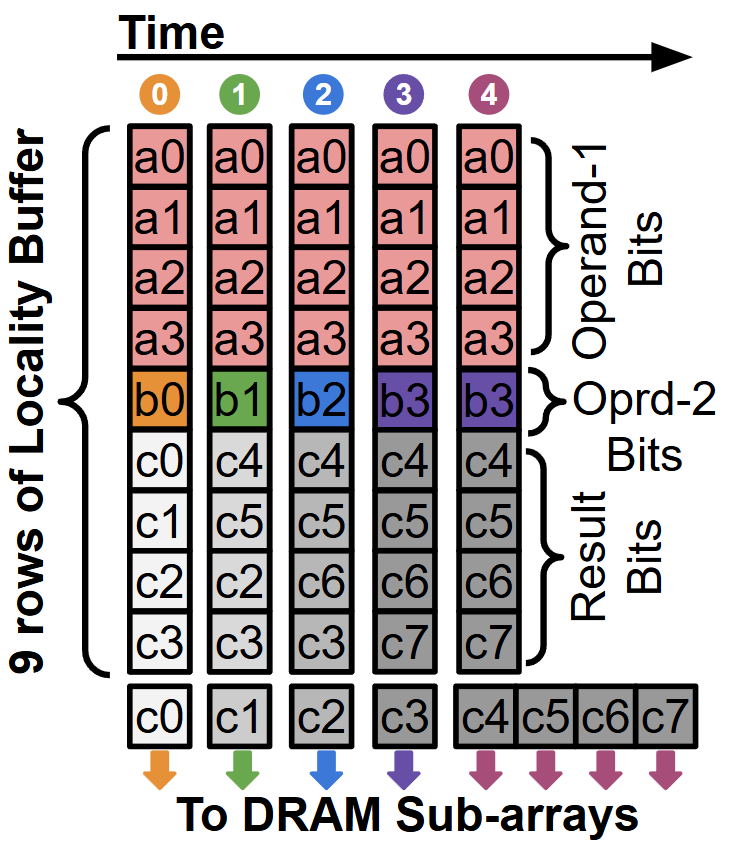}
        \caption{Data layout in 1 column of \textit{locality buffer} across steps. Same for all columns.}
        \label{fig:bits_row_buffer}
     \end{subfigure}
     \caption{Example of int4 bit-serial multiplication.}
     \label{fig:compute_scheme}
\end{figure}

\subsection{System Overview}

% The \sysname system overview is illustrated in Fig.~\ref{sys_overview}; detailed DRAM architectural elements are omitted. 
% The system interfaces with the host processor via a standard DRAM controller and operates as conventional DRAM in \textit{storage mode}. When the host processor initiates \textit{compute mode}, an embedded microcontroller within the DRAM controller issues specialized PIM instructions to enable in-memory computation.

\sysname~adds extra units on conventional DRAM and modify the host-DRAM interface to perform efficient reuse-aware in-DRAM processing. As illustrated in Fig.~\ref{sys_overview}, \sysname ~employs 1) extra computing units including locality buffers, bit-serial Processing Elements (PEs) and Popcount Reduction Units at each bank, 2) extra control units - a finite state machine (FSM) at each chip(device), shared by all banks. and 3) extended PIM command interface for the host processor, summarized in Table. \ref{tab:pim_cmds} . 
The DRAM controller on host processor is enhanced to send PIM commands through the DRAM command interface. The pim\_enable command toggle the DRAM into PIM mode where all incoming PIM commands are decoded by the FSM. 
For \texttt{broadcast\_enable, broadcast\_disable} commands, FSM simply configures the Mode Register Set (MRS) and set the data path control signals. 
For \texttt{pim\_add, pim\_mul, pim\_mul\_red, pim\_add\_parallel} commands, the FSM issues micro-ops to PEs, locality buffer, popcount units and subarrays to perform a corresponding computation. When a PIM kernel is completed, the pim\_disable command toggles off the PIM mode and results can be accessed through normal DRAM reads.
PIM commands are encoded with previously unused or vendor-reserved command encodings in the DRAM command/address protocol. The operand and control fields are transferred through address bus across multiple cycles depending on field length. 

% Using a matrix multiplication (matmul) workload as an example, the host processor first determines (or is provided with) an optimal mapping strategy and transmits it, along with workload specifications, to the microcontroller. This strategy defines the data layout for both the M-K and K-N matrices as well as internal broadcasting schemes. The microcontroller translates this strategy into a sequence of PIM instructions, starting with commands to load the K-N matrix into subarrays. It then activates the appropriate broadcasting units to load the M-K matrix according to the designated layout. Once the M-K matrix is loaded and broadcast to the target regions, a compute instruction is issued to perform the in-memory computation. 
% Computation is performed within the DRAM and the results are available in the subarrays upon completion of the requested instruction. The results are then written back to the host processor. For an end-to-end LLM workload, the LLM is first converted into multiple consecutive Matmul kernels, which are executed sequentially in PIM device. In our model, we assume that the weight matrices are already stored in the DRAM for each kernel.

%To couple with the flexible mapping strategies and have better data reuse in bit-serial scheme, 
In the following subsections, we describe each type of extra units added upon conventional DRAM.
% including bit-serial PE (Section \ref{bit_serial_PE_sec}), popcount reduction (Section \ref{popcount_adder_sec}), locality buffer (Section \ref{locality_buffer_bit_serial_multi}) and broadcasting units (Section \ref{broadcast_unit_sec}).

\subsection{Bit-Serial PE}
\label{bit_serial_PE_sec}
Fig.~\ref{bit_serial_pe} presents the schematic of the \textit{PE} attached to columns of the \textit{Locality Buffer}. When \textit{B} is positive, the PE performs a 1-bit full add with carry for \textit{C} and \textit{A}. When \textit{B} is negative, the PE routes \textit{C} to \textit{Output} and do not update the carry bit. 
To perform a bit-serial addition of operands \textit{op1} and \textit{op2}, \textit{B} is set to positive. At each cycle, one bit of \textit{op1} and \textit{op2} are sent to \textit{Result} and \textit{A} , and one bit of \textit{C} is populated at \textit{Output}.
For bit-serial multiplication, each cycle the PE update one result bit, depending on the current result bit and corresponding \textit{op1} and \textit{op2} bit values. At each cycle, one bit of \textit{op1} is sent to \textit{A}, one bit of \textit{op2} is sent to \textit{B}, and one bit of result is sent to \textit{C}. If \textit{op2} bit is 0, there is no need to update the result. If \textit{op2} bit is 1, the result bit is updated by adding with the op1 bit, and the carry bit is updated.

% The PE reads two 1-bit operands (A,B) from the DRAM cells and one input carry bit from the carry register. Operands A and B are processed by the \textit{SGEN} block to generate the sum bit and by the \textit{CGEN} block to compute the carry bit. The output (\textit{Product} or \textit{sum}) is selected by \textit{sel}, depending on the current operation mode. The output carry bit is written back if the current PE is processing the MSB bit in addition mode. 

\subsection{Locality Buffer \& Bit-serial Multiplication}
\label{locality_buffer_bit_serial_multi}

To facilitate bit-level reuse, a \textit{Locality Buffer} is added to each bank to accommodate full operand reuse for bit-serial multiplication. Fig.\ref{fig:compute_scheme} shows an example of int4 bit-serial multiplication using the \textit{locality buffer}.

Fig.~\ref{fig:bit_serial_compute} illustrates the 5 computing steps of the int4 bit-serial multiplication and \ref{fig:bits_row_buffer} shows contents of in locality buffer rows at each step. 
\circled{0} bit\#0\textasciitilde3 of operand-1 and bit\#0 of operand-2 are loaded to 5 rows of the buffer. Result bit\#0\textasciitilde3 are equal to operand-1 If the bit\#0 is 1, otherwise 0s. Result bit\#0 is then populated back to DRAM sub-array. \circled{1} Load bit\#1 of operand-2, if that bit is 1, serially update result bits\#1\textasciitilde5 by adding to operand-1 bitss\#0\textasciitilde3. Bit\#1 is immediately populated back to DRAM after updated. \circled{2} Similarly, load operand 2 bit\#2 to row buffer, update results bits\#2\textasciitilde6 and populate bit\#2. \circled{3} Similarly, result bits\#3\textasciitilde7 are updated in row buffer and result bit\#3 is populated. \circled{4} There is no more update on result bits. Bits \#4\textasciitilde7 are serially populated. 

Using this compute scheme, each operand bit is loaded into the row buffer only once, and each result bit is written to the DRAM array only once. This compute scheme reduces number of DRAM row accesses needed to perform a n-bit multiplication from $O(n^2)$ to $O(n)$ which greatly improves the efficiency of multiplications.
% However, this scheme introduce more row buffer area as a performance-cost tradeoff. 
To accommodate full reuse of $n$-bit integer multiplication, $2n+1$ rows are required. We select 17 rows of locality buffer for full reuse of up to 8-bit integer multiplications.

\sysname ~leverages SALP-MASA\cite{salp} mechanism to provide highest bandwidth to locality buffer.  A row of sub-array is activated before it is accessed and kept activated or precharged while other subarrays are activated. This mechanism saturates the global bitline and provides the highest data bandwidth to locality buffer. To efficiently use this mechanism, rows to be accessed successively in a block are mapped to different sub-arrays to allow overlapped activation.

% Fig. \ref{sys_overview} shows the overview of \sysname system; detailed DRAM architecture is omitted. The PIM system is interacting with the host processor through two paths, the \textit{data path} for regular DRAM reads and writes, and the \textit{command path} specifically for PIM instructions. A micro-controller is added to the regular DRAM controller to generate PIM instructions. When receiving PIM instructions, the micro-controller activates the peripheral circuits to perform in-DRAM vector operations. Peripheral circuits include the broadcasting units (Section \ref{broadcast_unit_sec}), \textit{locality buffer} (Section \ref{locality_buffer_bit_serial_multi}), , \textit{bit-serial PEs}(Section \ref{bit_serial_pe_sec}), and popcount reduction (Section \ref{popcount_adder_sec}).

\subsection{Popcount Reduction}
\label{popcount_adder_sec}
Workloads like matrix-multiplication require cross-column reduction depending on mapping. \sysname~ leverage popcount reduction units added to each bank to perform efficient cross-column reduction. As the vertical data layout allowing only 1 bit of all operands to be accessed at each cycle, a popcount unit is more efficient comparing to a reduction tree. As shown in Fig.~\ref{popcount_adder}, the popcount reduction unit consists of a popcount module and an accumulator, and supports reduction across columns.
% Inside the DRAM array, each column has one partial sum in a transposed layout. In this example, each DRAM array supports up to 1024 16-bit partial sums. The reduction is performed across columns, starting from the MSB to LSB. The slice of MSB bits is first passed through the pop-count logic, which will output the number of 1's in the input slice. It is then further padded to 26-bit and accumulated with the 27-bit result buffer. For each bit-slice, the partial sum contributes to the reduction sum in a shift manner. 
At each cycle, the popcount module fetches 1 bit of all operand in columns (bit-slice) and calculates number of 1's as a integer. This integer is then shifted and accumulated to the sum.
The sum is updated as $sum=sum+popcount(bitslice_i)\cdot 2^i$. 
% In this example, the slice of MSB bits is shifted 16 times, while the last slice, the slice of LSB bits, does not require a shift after 16 clock cycles. 
The \texttt{pim\_mul\_add} command fuses a multiplication with column-wise reduction as they can be efficiently pipelined.
The reduction result is written back to DRAM arrays in a horizontal layout to save row activations.
In addition, the \texttt{pim\_add\_parallel} leverage the accumulator inside the popcount reduction unit to perform a fast int32 bit-parallel addition, which is useful to add up multiple reduction results.
% \textcolor{red}{TODO: add a 32-bit addition data path and re-calculate area }

\subsection{Broadcasting Units} 
\label{broadcast_unit_sec}
To efficiently parallelize workload, some data needs to be broadcasted across one or a few DRAM hierarchies. Although this can be achieved by explicitly write the same data, it introduces redundant data transfers and stresses the DRAM data bus. With hardware broadcasting units, those redundant data writes can be eliminated. As shown in Fig. \ref{broadcast_unit}, broadcasting units are added to the bank and column level to leverage the higher internal bandwidth.
At the bank level, each bit of input can be broadcast through a \textit{demux} from the datapath to all banks. When bank broadcasting is enabled, the 64-bit input is broadcast to multiple banks selected by the \textit{Bank Select} signal.
Similar to the bank broadcast, at the column level, a \textit{demux} can broadcast each bit to multiple columns of the global row buffer where data is further written back to sub-arrays. 
% When subarray broadcasting is enabled, the 64-bit input is broadcast to the specified columns, selected by \textit{column select} signal, of each subarray. It is possible to implement broadcasting units across different DRAM hierarchies; however, due to the complexity of DRAM interconnect circuitry, we only consider cross-bank and cross-subarray broadcasting in this paper.

%% file: tables/pim_cmd.tex
\begin{table*}
\centering
\scriptsize
\caption{Extended PIM Commands and Their Instruction Encodings.}
\vspace{-0.5em}
\begin{tabular}{@{}lcccl@{}}
\toprule
\textbf{Instruction} & \textbf{Opcode Field} &
\textbf{Operand Fields} & \textbf{Control Field} &
\textbf{Description} \\ 
\midrule
\texttt{pim\_enable} &
000010 & -- & -- &
\makecell[l]{Enables the PIM operating mode through a Mode Register Set (MRS) write} \\

\texttt{pim\_disable} &
000011 & -- & -- &
\makecell[l]{Disables PIM mode and restores normal DRAM command decoding} \\

\texttt{broadcast\_enable} &
000000 & -- & bank\_bc, col\_bc &
\makecell[l]{Enables broadcast write mode of bank or column} \\

\texttt{broadcast\_disable} &
000001 & -- & -- &
\makecell[l]{Disables broadcast mode} \\

\texttt{pim\_add} &
010000 & R\textsubscript{dst}, R\textsubscript{src1}, R\textsubscript{src2} &
prec[3:0] &
\makecell[l]{Performs bit-serial addition of operands R\textsubscript{src1} and R\textsubscript{src2}, storing the result in R\textsubscript{dst}.} \\

\texttt{pim\_mul} &
010001 & R\textsubscript{dst}, R\textsubscript{src1}, R\textsubscript{src2} &
prec[3:0] &
\makecell[l]{Performs bit-serial multiplication between R\textsubscript{src1} and R\textsubscript{src2}, storing product in R\textsubscript{dst}.} \\

\texttt{pim\_mul\_red} &
010010 & R\textsubscript{dst}, R\textsubscript{src1}, R\textsubscript{src2} &
prec[3:0] &
\makecell[l]{Performs bit-serial multiplication followed by column-wise popcount reduction.} \\

\texttt{pim\_add\_parallel} &
010011 & R\textsubscript{dst}, R\textsubscript{src1}, R\textsubscript{src2} &
- &
\makecell[l]{Performs bit-parallel addition using the adder inside the popcount reduction unit.} \\

\bottomrule
\end{tabular}
\vspace{-0.75em}
\label{tab:pim_cmds}
\end{table*}

%% file: sec/04_mapping.tex
\section{Workload Mapping}
\label{sec:mapping}
% \begin{figure*}
%      \centering
%      \begin{subfigure}{0.49\textwidth}
%          \includegraphics[width=\columnwidth]{figs/tiling.png}
%          \caption{Hierarchical Mapping\{\textit{M: RB, N: CD, K: A}\} of complete matmul ($M,K, N$). Cross-array reduction is performed by the host.}
%          \label{fig:spatil_mapping}
%      \end{subfigure}
%      \hfill
%      \centering
%      \begin{subfigure}{0.3\textwidth}
%          \centering
%          \includegraphics[width=\columnwidth]{figs/array_mapping.png}
%          \caption{Block Mapping\{\textit{$R:M_tN_t,\ C:K_t$}\} of tiled matmul ($M_t,K_t,N_t$). The popcount reduction unit is not shown.}
%          \label{fig:array_mapping}
%      \end{subfigure}
%      \caption{\textcolor{red}{Example of hierarchical-mapping and array-mapping strategy.}}
%      \label{fig:workload_mapping}
% \end{figure*}

\begin{figure*}
     \centering
     
         \includegraphics[width=\textwidth]{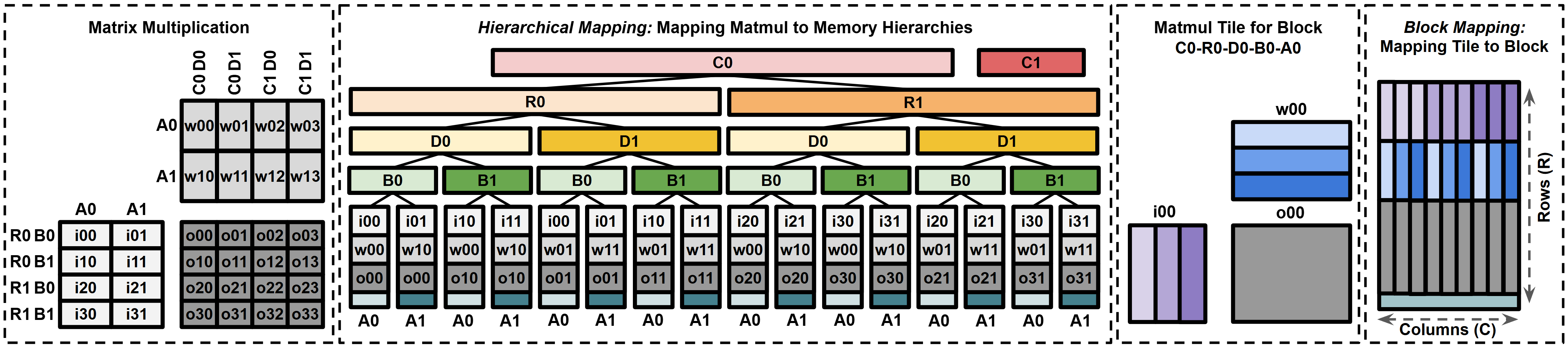}
         \caption{Example of Hierarchical Mapping \{\textit{M: RB, N: CD, K: A}\} and Block Mapping \{\textit{$R:MN,\ C:K$}\} of Matmul.}

     \label{fig:workload_mapping}
\end{figure*}

In order to accomodate flexible workload sizes and efficient exploration of workload mapping, we propose a mapping framework that supports flexible mapping of arbitrary-sized matrix–matrix multiplication (matmul) on arbitrary \sysname ~hardware configuration. 
% \textcolor{red}{TODO: since we now have 1024 PEs less than the subarray column width (16K), we have a logical view of arrays: n rows x 1024 cols, and we further divide it to partitions of nx256, as the unit for array mapping. Need to add a figure showing how logical arrays map to physical subarrays} 
In order to efficiently explore parallelism, the mapping framework views the sub-arrays of DRAM as many vertically-divided \textit{Blocks} as the sub-arrays are usually too wide to be mapped naively. A projection from \textit{block} to \textit{sub-array} is performed to determine the physical mapping. For parallelism levels C (\textit{Channel}), R (\textit{Rank}), B (\textit{Bank}), D (\textit{Device}), and A (\textit{Block}), we define \textit{Hierarchical Mapping} as a projection from each GEMM dimension to certain hierarchies. Within each \textit{block}, we define \textit{Block Mapping} as the projection of R (\textit{Row}) and C (\textit{Column}) to any GEMM dimensions.
For example, Hierarchical Mapping \{\textit{M: RB, N: CD, K: A}\} means the M dimension is hierarchically mapped to \textit{Ranks} and \textit{Banks}, N dimension is hierarchically mapped to \textit{Channels} and \textit{Devices}, and K dimension is mapped to \textit{Blocks}. Block Mapping \{\textit{$R:MN,\ C:K$}\} means \textit{rows} of a \textit{block} holds the M and N dimension, while \textit{columns} holds the K dimension. 
% Our mapping strategy consists of two components: \textit{Hierarchical Mapping}, which assigns matmul dimensions to parallelism dimensions above the subarray level, and \textit{Array Mapping}, which handles the mapping within subarrays.
% Note that, although additional peripheral circuits are integrated into standard DRAM to enable efficient computing in \sysname, our mapping strategy discussed here is general and  applies to other DRAM-PIM architectures as well. 

\subsection{Hierarchical Mapping}
Fig.~\ref{fig:workload_mapping} shows an example of mapping a matrix multiplication of size \textit{(M, K, N)} on DRAM with 2-channel, 2-rank, 2-bank, 2-device, and 2-block. Mapping on channel-1(C1) is not shown due to space limit. The matmul's M-dimension is hierarchically tiled to \textit{ranks} and \textit{banks}, the K-dimension is tiled to \textit{blocks} and the N-dimension is hierarchically tiled to \textit{channels} and \textit{devices}. The matmul tiles mapped to each block as well as the broadcasting patterns can be inferred by the C,R,D,B,A indices. Note that each input matrix tile is duplicated in 2 devices, indicating need to broadcast. A cross-block reduction can be implied as reduction dimension K is mapped to blocks.
% The 4 tiles in the figure are scheduled only to different subarrays but same index of channel, rank, bank, and device. tile-1 and tile-3 are scheduled to the $subarray_0$ of $(channel_1,rank_0,bank_1,device_1)$, while tile-2 and tile-4 are scheduled to the $subarray_1$. Moreover, tile-3 and tile-4 also contribute to the computation scheduled on $bank_0$. This sharing then notifies the simulator to enable the cross-bank broadcasting unit, as well as other broadcasting units if implemented. This \textit{hierarchical tiling} utilizes all the parallelisms of the DRAM hierarchy and also provides a simple indexing scheme to track the tiling mechanism and computation flow within the PIM device. 

\subsection{Block Mapping}
 Fig.~\ref{fig:workload_mapping} shows an example of block mapping \{\textit{R: $M_t$$N_t$, C: $K_t$}\} on the right side. After \textit{hierarchical mapping}, each block is assigned a tile, and the \textit{block mapping} further determines the block data layout and computation scheme. In this example, each column of the block  contains one slice of matrix $i00$, one slice of matrix $w00$, and computes a partial sum of matrix $o00$. A column-wise reduction is performed across columns. The SIMD multiplications and the column-wise reduction are fused as a \texttt{pim\_mul\_red} instruction for efficient bit-level pipelining.
 % The computation occurs along the K-dimension, which means the bit-serial operation happens across the rows in a SIMD fashion. 
 
 % The data layout is significantly different than normal DRAM data layout, and may be even more complicated in some array-mapping strategies. We assume the data layout reorganization is performed by the host, and the latency overhead is dominated by the DRAM I/O latency. 
 
% Once \textit{Hierarchical Mapping} and \textit{Block Mapping} are determined, data layout, broadcasting and reduction can all be inferred. For example, in the previous hierarchical mapping \{\textit{M: RB, N: CD, K: A}\}, the \textit{MK} matrix has rows tiled evenly across ranks and banks, columns tiled across subarrays, and the whole matrix is duplicated in all channels and devices, which requires broadcasting. \textit{KN} and \textit{MN} matrix layout can be similarly inferred. In the previous array mapping \{\textit{R: $M_tN_t$, C: $K_t$}\}, each subarray column contains one column of \textit{$M_tK_t$} matrix tile, one row of \textit{$K_tN_t$} matrix tile, and the whole \textit{$M_tN_t$} matrix tile.
%Fig. \ref{fig:array_mapping} illustrates an example of array mapping.

% Reduction across any parallelism levels can also be inferred. Hierarchical mapping \{\textit{M: RB, N: CD, K: A}\} indicates cross-subarray reduction as \textit{K} dimension is mapped to subarrays, and array mapping \{\textit{R: $M_tN_t$, C: $K_t$}\} indicates further cross-column reduction as $K_t$ dimension is mapped to columns.

\subsection{Scheduling}
After tiling and mapping, if the block is able to hold the entire matmul tile (i.e. with \{\textit{R:$MN$, C:$K$}\}, $K_t\leq\#Cols$ and $M_t\times N_t\leq \#rows$), the tile is scheduled to the block without further temporal tiling. In the case that the matmul tile exceeds the size that a block can handle, a further \textit{temporal tiling} is performed to iteratively schedule smaller tiles on a block. For this mapping example, the number of iterations would be $\frac{M}{\sqrt{\#\text{rows}}} \times \frac{N}{\sqrt{\#\text{rows}}} \times \frac{K}{\#\text{cols}}$.

% \[
% \frac{M}{\sqrt{\#\text{rows}}} \times \sqrt{\#\text{blocks}} \times \#\text{cols}}.
% \]

\subsection{Mapping Framework}
\begin{figure}
    \centering
    \includegraphics[width=0.8\columnwidth]{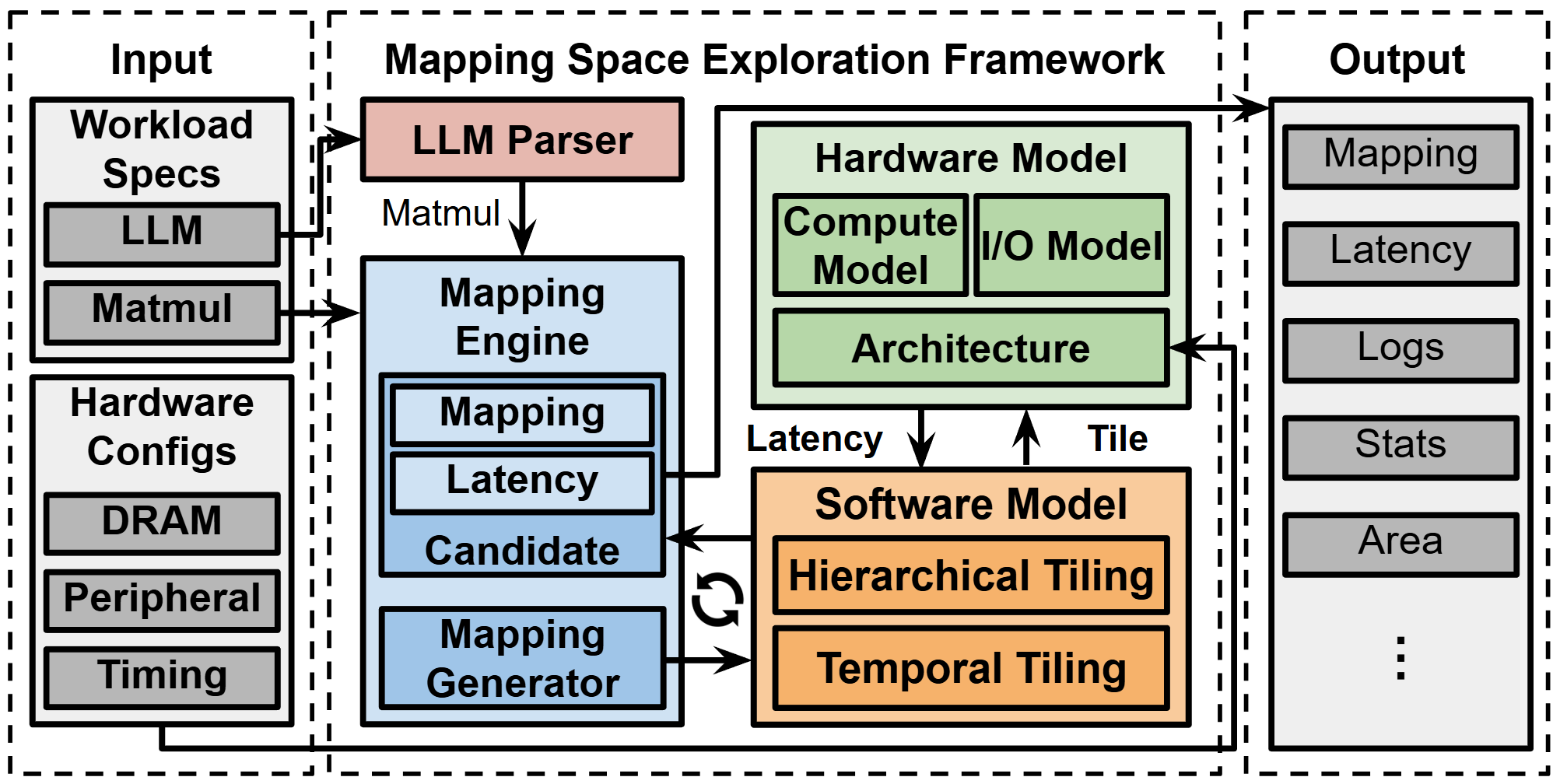}
    \caption{\sysname ~Mapping Framework}
    \label{map_framwork}
\end{figure}

Fig.~\ref{map_framwork} illustrates the \sysname~mapping space exploration framework. The framework takes as input the workload specifications (either LLM hyper-parameters or standalone matrix–matrix multiplication shapes) and the hardware configuration, including DRAM organization, peripheral-unit configuration, and timing parameters. The hardware model parameters are summarized in Table~\ref{params}. For each workload–hardware pair, the framework produces an optimized mapping together with the corresponding end-to-end latency, as well as auxiliary logs, statistics, and area-related reports.

The mapping space exploration framework is composed of an LLM parser, a mapping engine, a software model, and a hardware model. The LLM parser decomposes an input LLM into a sequence of matmul kernels and feeds them to the mapping engine. The mapping engine maintains the current best candidate mapping and its latency. A mapping generator inside the engine enumerates the mapping space, instantiates each mapping, and evaluates it using the software and hardware models, updating the candidate when a lower-latency mapping is found.

The software model applies hierarchical and temporal tiling according to a given mapping, performs scheduling across DRAM hierarchies, and issues per-tile compute and data-movement requests to the hardware model. For each tile, the hardware model returns the computation latency and the I/O latency. The software model accumulates these latencies across temporal tiles; the total kernel latency is the sum of all tile compute latencies and I/O latencies, and is returned to the mapping engine as the objective value of the current mapping.

The hardware model consists of an architectural description, a compute model, and an I/O model. The architecture is parameterized by the input hardware configuration. Given a tile and its mapping, the compute model computes the block-level PIM latency by summing the latencies of all PIM instructions executed on the locality buffers, PEs, and reduction units. The I/O model estimates the latency of interactions with the host CPU across DRAM hierarchies during input broadcasting and output reduction and collection, based on the inferred data layout, the amount of I/O traffic, the effective bandwidth, and the configuration of the broadcasting units. Together, these models enable \sysname\ to explore the full mapping space and identify latency-optimal mappings for each kernel.

\begin{table}[h!]
\centering
\caption{Hardware Model Parameters}
\begin{tabular}{lcccc}
\hline
\textbf{Type} & \textbf{Parameters} \\
\hline
DRAM Configuration & \makecell{\#Channels, \#Ranks, \#Banks, \#Devices,\\ \#Subarrays, \#Rows, \#Cols\\ Device Data Width, Frequency \\global bitline bus width} \\
\hline
Peripheral Units Configuration & \makecell{\#PEs, \#Locality Buffer Rows, \\Popcount Reduction Unit Width,\\ Broadcasting Units Width}   \\
\hline
Timing Parameters & \makecell{$T_{rcd}$, $T_{rp}$ \\ PE Latency, \\Locality Buffer Access Latency,\\ Popcount Reduction Latency} \\
\hline
\label{params}
\end{tabular}
\end{table}

%% file: sec/05_methodology.tex
\section{Methodology}
\label{sec:methodology}

\subsection{Performance Modelling}

Similar to previous works' approach\cite{optipim, llmcompass, calculon}, our hardware model is analytical, and hierarchically models the DRAM organization, broadcasting units, processing elements, locality buffers, reduction units and their interconnect. We validate the DRAM timing parameters and bandwidth model with Ramulator\cite{ramulator} and publicly available spec sheets. We implement the added peripheral units in Verilog\cite{verilog} and obtain their timing parameters using Synopsys Design Compiler\cite{synopsysdc}. Our LLM Parser is built on top of \textit{LLMCompass} \cite{llmcompass}

\input{sec/05_1_area_estimation}

\subsection{Workloads}
We evaluate \sysname ~on 4 end-to-end LLMs and their prefill and decode stages. Tabel \ref{tab:workloads} summarizes the models' layer, hidden size, number of heads, and quantization precisions. Following the same approach as prior works \cite{h2llm, dfx}, we choose 2 senarios for LLM end-to-end inference: \textit{Code Generation} with 1024 prompt tokens and 4096 output tokens as the "prefill heavy" scenario, and \textit{Context Understanding} with 8192 prompt tokens and 256 output tokens as the "decode heavy" scenario. For prefill stage throughput evaluation, we use 1024 as prompt token length. We use batch size=1 for all workloads. 

\subsection{Evaluated Systems}
Table \ref{sys_configs} shows the systems and configurations we used in evaluation. The H100 system contains an H100 GPU and 512GB host DRAM for data offloading. We use LLMCompass\cite{llmcompass}, a well-validated simulation framework to evaluate the H100 system latencies. The Proteus system is "consistent" with original paper which contains 1 Channel, 1 Rank and 16 Banks, and we assumes 512GB non-PIM host memory for data offloading. We use Proteus' open-source simulator to obtain its latencies. For both H100 and Proteus systems, the added host DRAM is to offload the LLM weights, and we assume zero offloading for those systems. The \sysname ~system configures all 1024GB host memory to PIM-enabled memory, and is evaluated using our open-source simulation framework.

\begin{table}[h!]
\centering
\caption{Evaluated LLM workloads}
\label{tab:workloads}
\renewcommand{\arraystretch}{1.15}
\setlength{\tabcolsep}{6pt}
\begin{tabular}{lccccc}
\toprule
\textbf{Model} & \textbf{Layers} & \textbf{Hidden Size} & \textbf{Heads} & \textbf{Quantization} \\
\midrule
GPT-3 6.7B & 32 & 4096 & 32 & int8 \\
GPT-3 175B & 96 & 12288 & 96 & int8 \\
Llama-3 8B & 32 & 4096 & 32 & int8 \\
Llama-3 70B  & 80 & 8192 & 64 & int8 \\
\bottomrule
\label {workloads}
\end{tabular}
\end{table}

% Table \ref{tab:workloads} shows matmul and LLM model inference workloads evaluated. Each LLM Model is seperated into prefill and decode workloads. For prefills, we use 1024 input tokens. For decode we use 2048 output tokens. For evaluation of LLM decode on \sysname, we apply layer pipelining where each layer is mapped to a seperate rank. This introduces a longer time to first token (TTFT) but facilitates higher throughput.

% \subsection{System Configurations}
\begin{table}[h!]
\centering
\vspace{-2mm}
\caption{System Configurations}
\begin{tabular}{lcccc}
\hline
\textbf{System} & \textbf{Configurations} & \textbf{TOPS}\tablefootnote{H100 TOPS is taken from \cite{h100whitepaper}. We calculate TOPS of Proteus and \sysname ~respectively based on their and our simulator.} \\
\hline
H100(PCIE)\cite{h100whitepaper} & \makecell{528 Tensor Cores,\\ 80GB HBM3 with 3352GBps bw \\512GB Offloading Memory\tablefootnote{NVIDIA reports 512GB LPDDR5X host memory for Grace Hopper Superchip: https://developer.nvidia.com/blog/nvidia-grace-hopper-superchip-architecture-in-depth/}} & 1978.9(int8) \\
\hline
Proteus\cite{proteus} & \makecell{1 out-of-order core \\ DDR5-5200, 1 Channel, \\ 1 Rank, 16 Banks \\ 512GB Offloading Memory} & 0.15(int8)  \\
\hline
\sysname & \makecell{1 out-of-order core \\ 1024GB DDR5, x16 \\8 Channels, 32 Ranks\tablefootnote{We envision future PIM-enabled memory systems to include more ranks than the JEDEC standard. Rambus already supports up to 16 ranks: https://www.rambus.com/memory-interface-chips/ddr5-dimm-chipset/ddr5-rcd/, and FB-DIMM, LR-DIMM\cite{multi-rank} effectively increase rank count to 64.}, 16 Banks,\\ 8 Devices, 128 Sub-arrays, \\128 rows, 16K columns per sub-array \\ 1024 PEs per Bank \\ 17x1024 Locality Buffer} & 986.9(int8)  \\
\hline
\label{sys_configs}
\end{tabular}
\end{table}

%% file: sec/05_1_area_estimation.tex
\subsection{Area Estimation}
\label{sec:area_estimation}
\subsubsection{DRAM \& locality buffer}
% Although DRAM area estimation tools have been proposed by academia, they fail to provide accurate area estimations due to the lack of detailed DRAM manufacturing parameters . This is particularly severe for DRAMs fabricated in advanced technology nodes such as DDR5 and HBM. To this end, w
We estimate the total DRAM area based on the reported single 16Gb DDR5 die area from \textit{Micron}\cite{micro_ddr5_area} instead of using DRAM area estimation tools \cite{DRAMSpec, CACTI7} due to the lack of detailed DRAM manufacturing parameters. We assume the DRAM chip storage density (area per bit) remains consistent across different configurations, and the total DRAM chip area is calculated by the product of the unit area per bit and the total storage bits.

The \textit{locality buffer} is modeled similar to aforementioned DRAM area calculation but using the bit-density reported by TSMC 45nm SRAM technology\cite{tsmc_45nm_sram}. This SRAM area is later added together with the peripheral logics as the final total added peripheral area. 

\subsubsection{Peripheral logic}
In practice, the periphery logics are fabricated using older, more mature process nodes to ensure greater reliability and thermal stability \cite{imec_dram_2020, spessot_dram_2015, edn_dram_2018}. Furthermore, DRAM peripheral circuits typically employ fewer interconnect layers than standard CMOS technologies \cite{date_dram_model_2013}. 

To evaluate the added peripheral area overhead, we first obtain the synthesis area from \textit{FreePDK 45nm} and \textit{Design Compiler}\cite{freepdk_45, design_compiler_manual}, and scale it down to 14nm, which is one generation older than modern DDR5 manufacturing tech node. Due the limited access to DRAM and SRAM physical design specs, we estimate the post-synthesis area based on some existing models and projections \cite{SpindlerDATE2007,probe,KahngPROBE2022}.There are three major factors affecting the post-synthesis area: placement utilization $U$, which reflects the area amplified due to place \& route; buffer growth factor $\beta \ge 0$, representing the area overhead from clock tree synthesis, timing repair, and resizing; and routing capacity $C$, which is primarily influenced by the number of metal layers used.The detailed derivation of these parameters is beyond this paper's scope and is omitted for simplicity. With these parameters and synthesis reports, the final post-synthesis area can be calculated.

%% file: sec/06_results.tex
\section{Results}
\label{sec:results}
% \subsection{Workloads}

% \multicolumn{4}{c}{\textbf{GEMM and GEMV Workloads}} \\
% \midrule
% \textbf{Matmul} & \textbf{M} & \textbf{K} & \textbf{N} \\
% \midrule
% GEMM\_small & 1024 & 12288 & 12288 \\
% GEMM\_large & 2048 & 24576 & 24576 \\
% GEMV\_small & 1 & 12288 & 12288 \\
% GEMV\_large & 1 & 24576 & 24576 \\
% \midrule
% \multicolumn{4}{c}{\textbf{LLM Workloads}} \\
% \midrule

% Table \ref{sys_configs} shows the system configurations of the evaluated systems. To accommodate LLMs, the H100 system has 512GB host memory\cite{h100_sys} and Proteus system has 512GB offloading memory. We assume zero offloading overheads for H100 and Proteus systems. The \sysname system has 1 out-of-order core for instruction offloading and data reorganization. The 512GB DDR5 main memory is enhanced with \sysname broadcasting units, locality buffer, and popcount reduction units.

In this section, we evaluate \sysname performance using several metrics: \textit{end-to-end throughput}, \textit{prefill throughput}, \textit{decode throughput}, and \textit{performance per $mm^2$}. We compare \sysname~against the NVIDIA H100 GPU and \textit{Proteus} (Section~\ref{sec:performance}). We then conduct an architecture ablation study (Section~\ref{sec:ablation}) and a sensitivity analysis (Section~\ref{sec:sensitivity}) to further characterize the sources of performance improvement and identify key design trade-offs.

% Table~\ref{tab:workloads} summarizes the evaluated workloads, including GPT-3 and Llama-3 LLMs. 
% For LLM inference, we separately evaluate the prefill and decoding stages, using token lengths of 1024 and 2048, respectively. For LLM decoding on \sysname, we employ layer pipelining, in which each layer is mapped to a separate DRAM rank. While this approach increases the time to first token (TTFT), it enables higher overall throughput. The complete system configurations used for evaluation are detailed in Table~\ref{sys_configs}.

\subsection{Performance}
\label{sec:performance}
% The normalized LLM end-to-end throughputs (also know as request throughput\cite{vllm}) of H100,  \textit{Proteus} and \sysname~ are shown in Fig.~\ref{fig:end-to-end-throughput}. \sysname~ achieves 90.1\texttimes~ and 15.6\texttimes ~higher geomean throughput than H100 on \textit{Code Generation} and \textit{Context Understanding} senarios respectively, while Proteus under-performs the H100 by orders of magnitude.

% Fig. ~\ref{speedup} shows the standalone prefill and decode stage throughputs of LLMs. Although prefill is generally considered a compute-bound task requiring high computational capabilities, where conventional PIM architectures often fall short, \sysname ~still achieves maximum 1.9\texttimes ~speedup on prefill stage. For the memory-bound decode stage, \sysname~ achieves maximum 112.0\texttimes ~speedup comparing to H100 due to high internal bandwidth and no need to transfer the static weights out of memory. Proteus achieve better relative performance in LLM decode stages than in prefill stages but still under-performs the H100.

% Fig. ~\ref{perf_per_area} shows the performance per area normalized to H100 on LLM inference workloads. \sysname ~demonstrates maximum 466.8\texttimes~ higher performance per area than H100 due to compact designs of peripheral units. Proteus shows better performance per area comparing to H100, but the gain is limited by inefficient multiplications without bit-level reuse.

\begin{figure}
    \centering
    % --- First subfigure ---
    \begin{subfigure}[t]{0.4\columnwidth}  % [t] aligns top
        \centering
        \includegraphics[width=\linewidth]{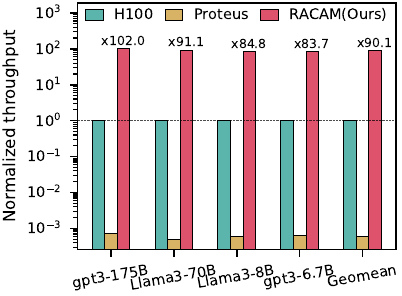}
        \caption{
        %End-to-end normalized throughput of 
        \textit{Code Generation} scenario (1024 input tokens, 4096 output tokens).}
        \label{fig:long_text_generation}
    \end{subfigure}
    \hfill
    % % --- Second subfigure ---
    % \begin{subfigure}[t] {0.48\textwidth} % [t] aligns top
    %     \centering
    %     \includegraphics[width=\linewidth]{figs/regular_chat_normalized_throughput.png}
    %     \caption{End-to-end normalized throughput of Regular Chat with 1024 input token and 128 output tokens.}
    %     \label{fig:regular_chat}
    % \end{subfigure}
    % --- Third subfigure ---
    \begin{subfigure}[t] {0.4\columnwidth} % [t] aligns top
        \centering
        \includegraphics[width=\linewidth]{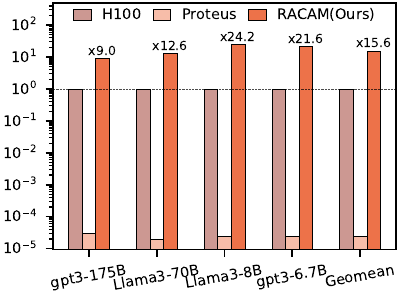}
        \caption{
        %End-to-end normalized throughput of 
    \textit{Context Understanding} scenario (8192 input tokens, 256 output tokens).}
        \label{fig:large_read}
    \end{subfigure}
    \caption{End-to-end normalized throughput of H100, \textit{Proteus} and \sysname ~in \textit{Code Generation} and \textit{Context Understanding} cases across evaluated gpt and Llama LLM models.}
    \label{fig:end-to-end-throughput}
\end{figure}

The normalized end-to-end LLM throughput (also referred to as \textit{request throughput}~\cite{vllm}) of H100, \textit{Proteus}, and \sysname~ is shown in Fig.~\ref{fig:end-to-end-throughput}. \sysname~ delivers \textbf{90.1\texttimes} and \textbf{15.6\texttimes} higher geometric-mean throughput than H100 on the \textit{Code Generation} and \textit{Context Understanding} scenarios, respectively, while \textit{Proteus} underperforms H100 by orders of magnitude.

\begin{figure}
\centering
    \includegraphics[width=0.6\columnwidth]{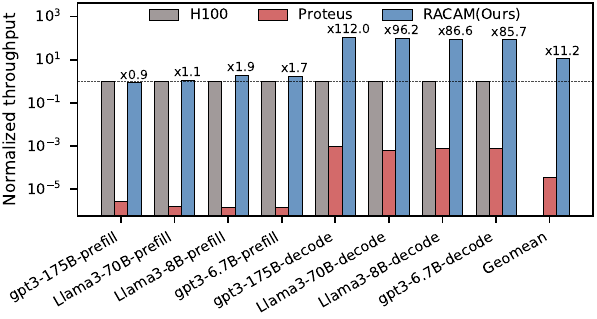}
    % \vspace{-0.3in}
    \caption{ Normalized end-to-end throughput of \textit{H100}, \textit{Proteus}, and \sysname.}
    \label{speedup}
\end{figure}

Fig.~\ref{speedup} presents the standalone prefill and decode throughput of LLM inference. Although prefill is widely regarded as a compute-bound phase---where conventional PIM architectures typically struggle---\sysname~ still achieves up to \textbf{1.9\texttimes} speedup over H100. For the memory-bound decode phase, \sysname~ reaches up to \textbf{112.0\texttimes} speedup due to its high internal bandwidth and the elimination of weight movement from DRAM. While \textit{Proteus} attains relatively better performance during decode than prefill, it still falls short of H100.

\begin{figure}
\centering
    \includegraphics[width= 0.6\columnwidth]{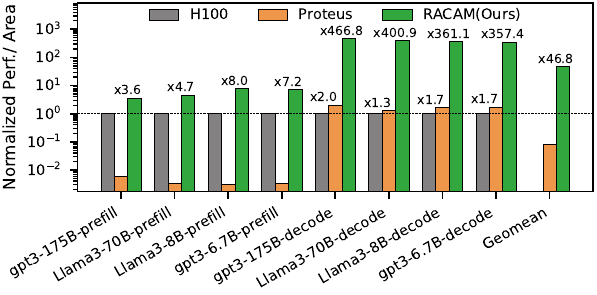}
    % \vspace{-0.3in}
    \caption{Performance per unit area of \textit{Proteus} and \sysname ~both normalized to performance per $mm^2$ of H100 GPU.}
    \label{perf_per_area}
\end{figure}

Fig.~\ref{perf_per_area} reports the performance per area normalized to H100.  We scale H100, Protues and \sysname area all to 15nm technology. \sysname~ incurs 4\% memory chip area overhead, and the total area of peripheral units is 24\% of the scaled H100 area\footnote{We calculate the H100 area as die + HBM, where HBM is flattened to 1 layer. Both die and HBM are scaled to 15nm tech node.}. We use 1\% to calculate area of added circuitry of Proteus as reported\cite{proteus, ambit}. \sysname~ achieves up to \textbf{466.8\texttimes} and \textbf{8.0\texttimes} higher performance per area in decode and prefill phases respectively owing to its compact peripheral designs. \textit{Proteus} also exceeds H100 in performance per area, but its improvement is constrained by inefficient bit-serial multiplications without bit-level reuse.

% \sysname achieves 1.3× higher overall throughput than the H100 on LLM prefill workloads. Although prefill is generally considered a compute-bound task requiring high computational capabilities, where conventional PIM architectures often fall short, \sysname delivers competitive performance compared to the H100. For smaller model sizes, when the prefill phase becomes less compute-bound, the advantages of high internal bandwidth and efficient data movement become more pronounced, leading to even higher throughput. For decode workloads, which are memory-bound, the combination of optimized computation mapping, high internal bandwidth, and minimal data movement results in a significant performance gain. Notably, the speedup scales with model size, further highlighting \sysname’s advantage in accelerating large-scale, memory-intensive workloads. Overall, favored by the high throughput of decode, end-to-end throughput of LLM workloads is further pushed to 11.2x higher on average.


% Fig.~\ref{perf_per_area} shows the performance per area normalized to H100 on LLM inference workloads. Overall, \sysname shows 77.6x higher performance density than H100.  

\subsection{Architecture Ablation Study}
\label{sec:ablation}
\begin{figure}
\centering
    \includegraphics[width= 0.6\columnwidth]{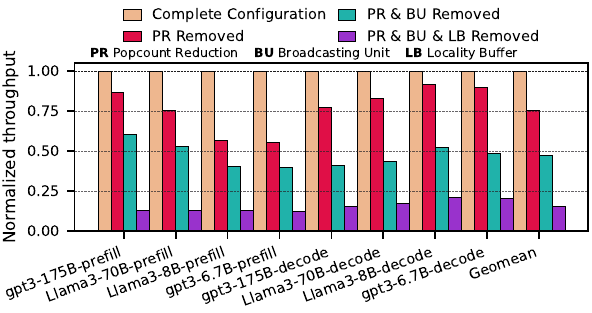}
    % \vspace{-0.3in}
    \caption{\sysname ~Ablation Study - performance is normalized to  configuration with all 3 enhancements (locality buffer, popcount and broadcasting unit).}
    \label{ablation}
\end{figure}

% \subsubsection{Architecture Ablation Study}
\label{sec:arch_ablation}

Fig.~12 presents an architectural ablation of \sysname{} across eight LLM workloads, where we progressively disable three key in-DRAM structures: the popcount reduction (PR) units, the broadcasting units (BU), and the locality buffers (LB). We normalize all results to the complete \sysname{} configuration and report the resulting end-to-end latency degradation. Removing only the PR units already increases prefill latency by $1.2$--$1.8\times$, and decode latency by $1.1$--$1.3\times$. This reflects the cost of exporting partial sums out of the subarray for off-array reduction at host CPU, which increases host data transfer and stresses external bandwidth instead of performing accumulation entirely within the local popcount reduction units.

When BU is also removed, latency roughly doubles relative to the full design for most decode workloads. Without BU, host CPU must duplicate input across DRAM banks and columns, increasing the host-DRAM data transfers. Removing BU generally has more significant impact on decode workloads, as they are less memory-bound than prefill workloads and more sensitive to data transfer latencies of kernel input.

Finally, eliminating the LB has the largest impact, because it forces all bit-level reuse to go back to the DRAM cell array. Prefill latency increases by about $7.5$--$8\times$ across all models, while decode latency increases by $4.7$--$6.5\times$. With LB, \sysname{} can keep the multiplicant bit near the PEs and exploit bit-serial reuse at row-buffer granularity; without LB, repeated ACT/PRE and global-bus transfers dominate execution time, shifting the design to strongly memory-bound. Overall, the ablation shows that all three structures---PR, BU, and especially LB---are essential to convert DRAM’s raw internal bandwidth into sustained, high-utilization in-DRAM compute for both prefill and decode.

\subsection{Sensitivity Study}
\label{sec:sensitivity}

\subsubsection{\textbf{PE Number Sensitivity}}
\begin{figure}
\centering
    \includegraphics[width= 0.6\columnwidth]{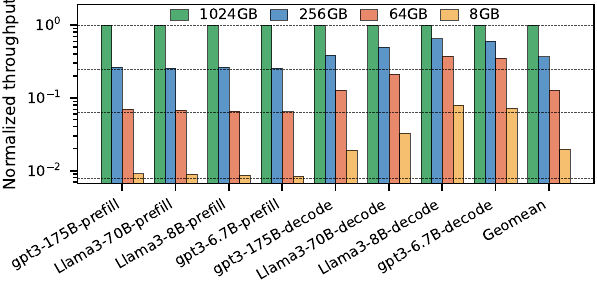}
    % \vspace{-0.3in}
    \caption{\sysname ~performance sensitivity to system capacity. Horizontal lines represents the normalized capacity ratios (1/8, 1/16, and 1/128), serving as reference for ideal performance scaling.}
    \label{sensitivity_pe_num}
\end{figure}
% Fig.~\ref{sensitivity_pe_num} presents a sensitivity study on total number of PEs in the system. Performance is evaluated as total number of $1/4$, $1/16$, and $1/64$ by reducing the number of channels and ranks.  LLM prefill workloads show almost linear scaling properties that closely align with the PE number shrinking reference as they being compute-bound. On the other hand, LLM decode workloads is less impacted by reduction in PE number and exhibit weak scaling behavior. LLM decode workloads demonstrate only modest performance degradation relative to the reference lines. This is because 1) they are less memory bound, and 2) their PE utilizations are relatively low due to their smaller problem sizes, making them less sensitive to variations in compute capacity. In particular, decoding kernels of smaller LLMs experience significantly less performance loss under reduced capacity as smaller LLMs' decoding kernels has lower PE utilizations, making them more invarient agaisnt compute capability reduction. 

Fig.~\ref{sensitivity_pe_num} presents a sensitivity study on the total number of PEs in the system. 
We evaluate performance under effective PE counts of $1/4$, $1/16$, and $1/64$ of the baseline by correspondingly reducing the number of channels and ranks. 
LLM prefill workloads exhibit near-linear degradation that closely follows the PE-reduction reference lines, as they are compute-bound and scale nearly proportionally with available compute resources.

In contrast, LLM decode workloads are far less affected by reductions in PE count and show weak-scaling behavior. Their performance decreases only modestly relative to the reference lines. 
This behavior arises because (1) decode is primarily memory-bound, and (2) typical decode kernels operate at relatively low PE utilization due to smaller per-token compute footprints, making them less sensitive to compute capacity.

Notably, smaller LLMs experience even less performance loss under reduced PE configurations, since their decoding kernels naturally have lower PE utilization and thus remain more invariant to reductions in compute capability.

% Compared to GEMV, LLM decode workloads, despite being memory-bound, degrade more gracefully. For example, the performance of Llama-3.1-8B decoding on an 8GB system remains above the 64GB reference scaling line. This is due to two factors. First, they often operate on much larger problem sizes than GEMV, which have better temporal locality. Second, \sysname ~employs a runtime mapping engine that assigns an optimal data layout to each LLM layer. This dynamic mapping improves operand reuse and minimizes unnecessary data movement, effectively mitigating the performance impact of limited memory capacity.

\subsubsection{\textbf{Precision Sensitivity}}
\begin{figure}
\centering
    \includegraphics[width= 0.6\columnwidth]{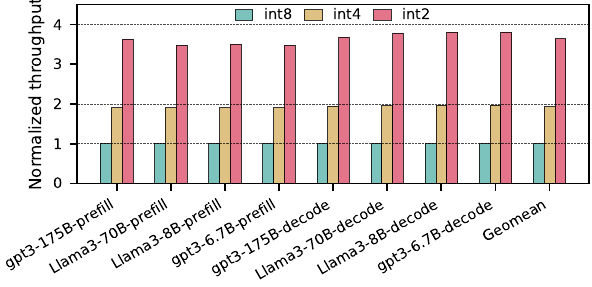}
    % \vspace{-0.3in}
    \caption{\sysname ~performance sensitivity to workload precision. \sysname ~uses bit-serial scheme, whose latency is ideally linear to data precision.}
    \label{sensitivity_precision}
\end{figure}
% Fig.~\ref{sensitivity_precision} presents a sensitivity study on workload precision, where data precision is varied from int8 (baseline) to int4 and int2 with expected performance increase of 2x and 4x. Overall, performance improves significantly along reduced precision across all workloads. When moving from int8 to int4, performance improves by approximately 2×, and further increases to around 3.5×–3.8× at int2, depending on the workload. The nearly linear scaling is directly led by the bit-serial compute with bit-reuse so that latency is linear to the operand bit-width for both addition and multiplication. Exactly linear scaling is not achieved due to the fixed (\texttt{pim\_add\_parallel}) latency incurred by the bit-parallel reduction across temporal tiles.

Fig.~\ref{sensitivity_precision} presents a sensitivity study on workload precision, where data precision is varied from \texttt{int8} (baseline) to \texttt{int4} and \texttt{int2}. 
Across all workloads, performance increases substantially as precision decreases. 
Reducing precision from \texttt{int8} to \texttt{int4} yields approximately 2$\times$ speedup, and further reducing to \texttt{int2} provides 3.5$\times$--3.8$\times$ improvement depending on the workload.

The near-linear scaling arises from the bit-serial compute design with bit-level reuse, which makes latency proportional to operand bit-width for both addition and multiplication. 
Perfect linear scaling is not achieved due to the fixed latency of the bit-parallel reduction (\texttt{pim\_add\_parallel}), which introduces a constant overhead that becomes more prominent at lower precisions.

%This demonstrates that the system effectively exploits the bit-serial compute model, especially at low precisions.

% Among all workloads, smaller LLM prefill workloads, gpt3-6.7b and Llama3.1-8b, benefit the most from reduced precision, showing ideal scaling.  The GEMV and LLM decode workloads also show significant improvements, although the gains are slightly sub-linear in some cases. This is likely due to their memory-bound nature. At int2, these workloads still benefit from reduced data size, which lowers I/O bandwidth pressure; however, but the effect is partially masked by other bottlenecks in the memory system.

\subsubsection{\textbf{Mapping sensitivity}}

\begin{figure}
\centering
    \includegraphics[width= 0.6\columnwidth]{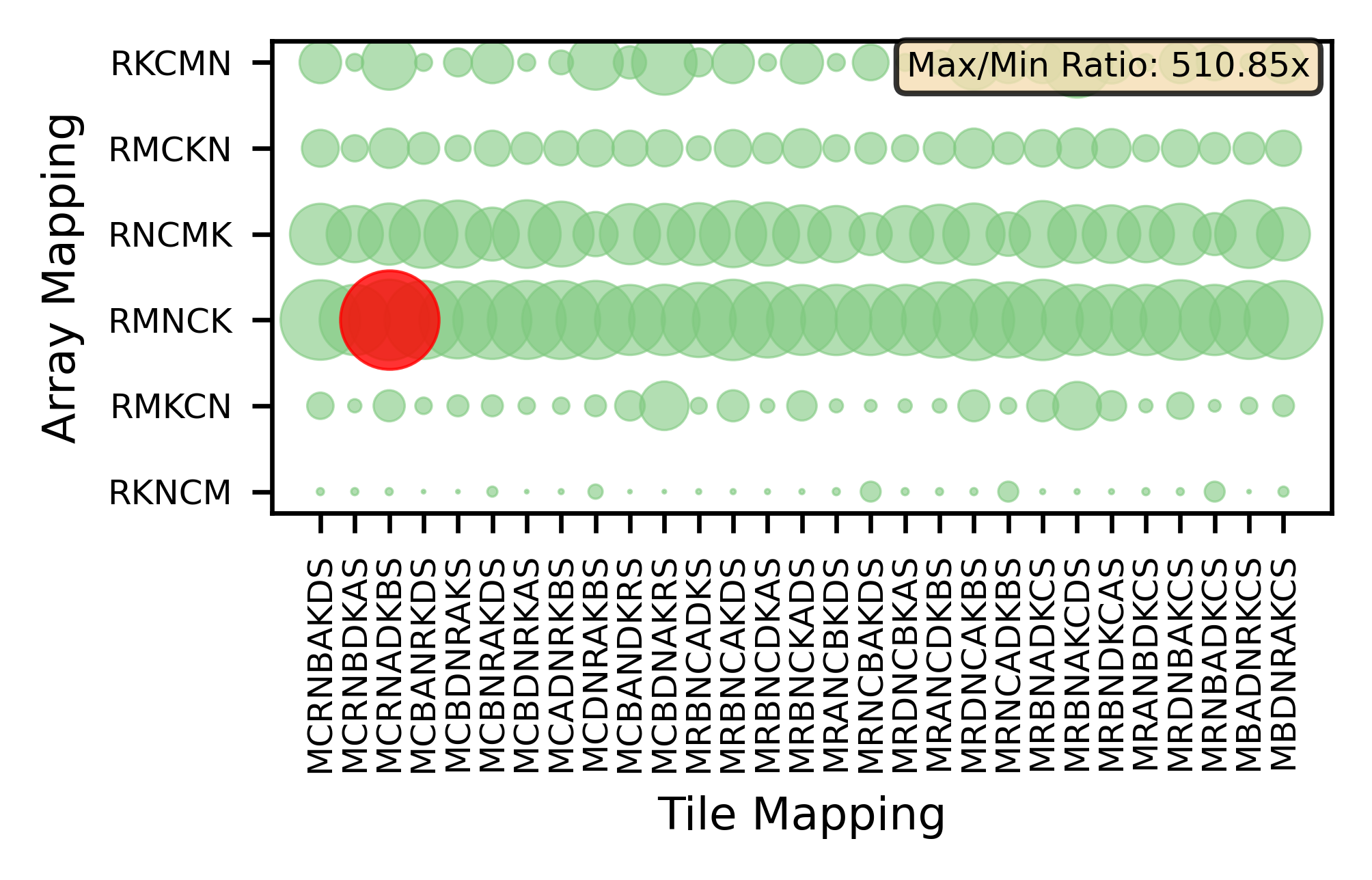}
    % \vspace{-0.3in}
    \caption{\sysname ~performance sensitivity to mapping strategies on 1024x12288x12288 GEMM. Larger circle means higher performance, red circle has the highest performance.}
    \label{sensitivity_mapping}
\end{figure}
% Fig.~\ref{sensitivity_mapping} shows the performance sensitivity to workload mapping on $1024\times12288\times12288$ GEMM size. The scatter plot demonstrates that both tile mapping and array mapping have a significant impact on the given workload, where the performance variance among mapping strategies can be very large (510.85\texttimes ~Max/min ratio). $RNCMK$ mapping shows significantly higher performance than other array mappings as it can leverage the popcount reduction unit to perform reduction across columns. Spatial Mapping shows unpredictable performance patterns signifying the importance of design-space search.

Fig.~\ref{sensitivity_mapping} shows the performance sensitivity to workload mapping for a $1024\times12288\times12288$ GEMM. 
The scatter plot illustrates that both tile mapping and array mapping exert substantial influence on performance, leading to large variability across mapping strategies (with a maximum--to--minimum ratio of 510.85$\times$). 
Among the array mappings, \texttt{RNCMK} achieves notably higher performance than others, as it can exploit the popcount reduction unit to efficiently perform column-wise reductions. 
In contrast, spatial mappings exhibit irregular and unpredictable performance trends, underscoring the necessity of systematic design-space exploration rather than relying on manually crafted mapping choices.

\subsubsection{\textbf{GEMM/GEMV Size Sensitivity}}
\label{sec:size_sensitivity}

Fig.~16 illustrates how GEMM and GEMV latency scales with problem size and how these trends correlate with PE utilization. Across all GEMM configurations, \sysname~ sustains near-ideal scaling despite the rapid growth of arithmetic intensity. For example, increasing the GEMM size from $2048\times 2048\times 2048$ to $32768\times 32768\times 32768$ enlarges the total compute requirement by $4096\times$, yet latency rises by only $2984.6\times$. This deviation from the capacity-scaled baseline corresponds directly to the increased PE utilization, which increases from $86.3\%$ to $98.0\%$. Because GEMM kernels achieve full utilization across columns, arrays, banks, channels, and devices (all at $1.0$), the improvement stems primarily from two factors: the ability of locality buffers to amortize operand loading as $K$ grows, and the dominance of compute latency over I/O latency at larger scales. For instance, in the largest GEMM, compute latency dominates total latency by $>98\%$, while I/O contributes only $1.39$ms out of $69.9$ms.

In contrast, GEMV kernels exhibit a different microarchitectural behavior, as shown in Fig.~16b. GEMV is inherently memory-bound, and thus the achievable PE utilization is much lower (e.g., $7\%$ for $1\times2048\times 2048$ GEMV compared to $82\%$ for large GEMMs). As GEMV size increases, PE utilization improves monotonically due to higher SIMD utilization and arithmetic intensity. This produces sub-linear latency growth: although workload size increases by up to $256\times$, observed latency increases by only $4\times$, demonstrating \sysname’s ability to convert additional operand reuse and hierarchical parallelism into effective speedup even for bandwidth-bound workloads.

% \subsubsection{GEMM/GEMV size sensitivity}

\begin{figure}
    \centering
    % --- First subfigure ---
    \begin{subfigure}[t]{0.49\textwidth}  % [t] aligns top
        \centering
        \includegraphics[width=\linewidth]{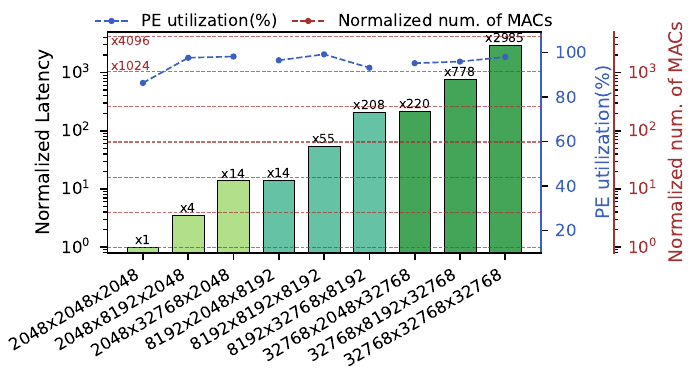}
        \caption{Sensitivity to GEMM sizes}
        \label{fig:sensitivity_gemm}
    \end{subfigure}
    \hfill
    % --- Second subfigure ---
    \begin{subfigure}[t] {0.47\textwidth} % [t] aligns top
        \centering
        \includegraphics[width=\linewidth]{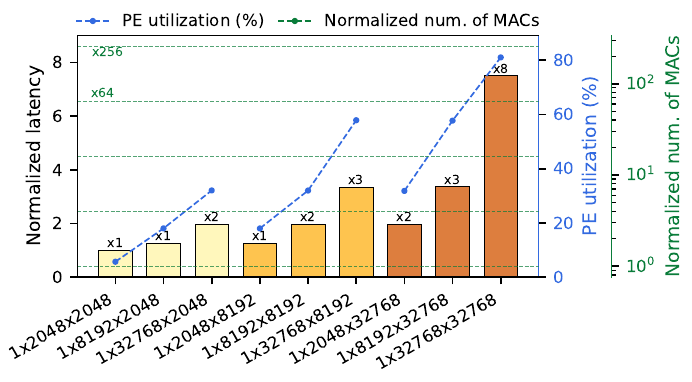}
        \caption{Sensitivity to GEMV sizes}
        \label{fig:sensitivity_gemv}
    \end{subfigure}
    \caption{Sensitivity analysis across GEMM and GEMV workloads noted in $M\times K\times N$. Sizes are grouped into 3, $M,N$ are fixed value and only $K$ is changing within each group.}
    \label{fig:sensitivity_matmul}
\end{figure}

\begin{figure}
\centering
    \begin{subfigure}[t]{0.3\columnwidth}
    \centering
        \includegraphics[width= \columnwidth]{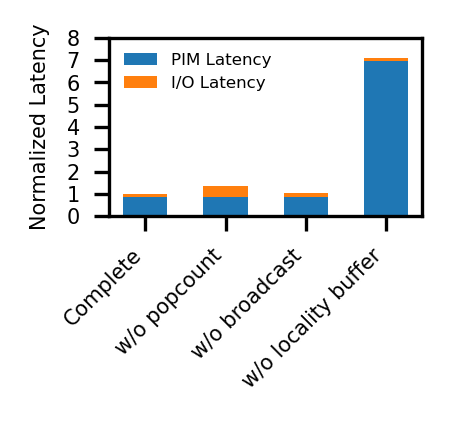}
        \caption{GEMM}
        \label{brkd-gemm}
    \end{subfigure}
    \begin{subfigure}[t]{0.3\columnwidth}
    \centering
        \includegraphics[width= \columnwidth]{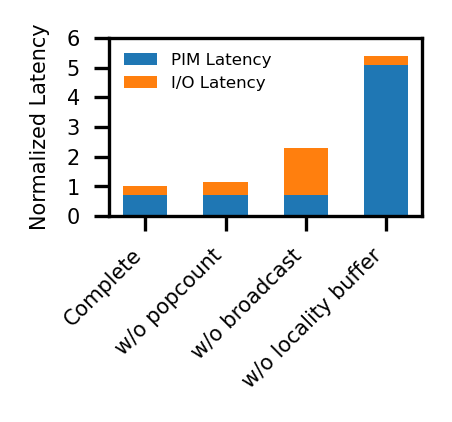}
        \caption{GEMV}
        \label{brkd-gemv}
    \end{subfigure}
    \caption{Breakdown of GEMM and GEMV latencies under ablation of hardware  as each component is removed.}
    \label{brkd}
\end{figure}

\subsubsection{\textbf{Latency Breakdown}}
\label{sec:breakdown}
Fig. \ref{brkd} illustrates the PIM and I/O latency breakdown of  a typical LLM kernel in prefill, GEMM-1x49152x12288, under different senarios of hardware unit ablation. 
“PIM Latency" refers to the total latency of PIM compute commands \texttt{pim\_add}, \texttt{pim\_mul}, \texttt{pim\_mul\_red}, \texttt{pim\_add\_parallel}. Larger blue portions indicate that the workload is more compute dominated.
"I/O Latency" refers to the total latency spend on communication with host, either to layout the kernel input data, fetch the kernel output data or perform host-side reduction. Larger orange portions indicate that the workload is more I/O dominated.

For the "Complete" case without ablation, GEMM latency is clearly compute dominated, corresponding to its compute-bound nature. Although GEMV is intrinsically memory-bound, \sysname ~leverages broadcasting units, high internal bandwidth, and the ability to store static weight data prior to computation to drastically reduce I/O latency.
Removing the popcount adder or broadcasting units both increase I/O latency as more data needs to go through host-dram I/O. 
Removing the locality buffer drastically increases PIM latency as no bit-level reuse can be exploited, and latency penalties are paid on repeated expensive ACT-PRE operations.

%% file: sec/07_discussion.tex
\section{Discussion}
\label{sec:discussion}

\textbf{Reliability:} Excessive concurrent row activations pose serious reliability risks in DRAM. Prior work shows that frequent, closely spaced activations accelerate charge leakage and can trigger disturbance errors in nearby rows, exhibiting RowHammer-like behavior~\cite{rowhammer}. This concern is amplified in PUD architectures, where massively parallel bit-serial operations require dense ACT--PRE sequences across many subarrays.

Unlike CPU-driven access patterns, which are naturally irregular and throttled by cache and instruction dependencies, PUD workloads generate highly regular activation patterns that repeatedly toggle the same wordlines, reducing the time available for cells to restore charge. As concurrent activations increase, vulnerable cells may exceed disturbance thresholds, imposing practical limits on how aggressively bit-level parallelism can be exploited. These challenges highlight the need for PIM architectures that reduce redundant ACT--PRE operations and schedule computation in ways that preserve DRAM integrity.

Moreover, prior studies show that even individual PUD operations are highly sensitive to timing, DRAM chip configuration, and vendor variation~\cite{computeDRAM}. AND/OR operations succeeded reliably on only 1 of 12 evaluated DRAM chips, further emphasizing the importance of using robust digital PEs rather than fragile charge-sharing mechanisms.

% \textbf{Time taken for Mapping Exploration:}
% Our exhaustive search mechanism of mapping can be finished within 1 second for a single GEMV, within 2\~3 seconds for a single GEMM, and within 5~10 seconds for LLMs on a 16-core consumer-level CPU. The search contains 1548 different mappings for GEMM and 192 for GEMV. GEMV has a smaller mapping space due to one dimension being 1 therefore no need to be tiled and mapped to any DRAM hierarchy. The exhaustive search can be completed quickly due to 1)Each search leverage an analytical performance model and can be completed in microseconds. 2) LLMs has consistent GEMM and GEMV kernel sizes across layers, therefore the same optimal mappings can be reused across layers.
% Furthermore, the search overhead can be amortized for end-to-end workloads like LLMs. The matrix dimensions are only affected by input token length during runtime, but as mapping is data-agnostic, mappings for different input token lengths can be pre-determined or cached during runtime to save runtime searching time.

\textbf{Time Taken for Mapping Exploration:}
Our exhaustive mapping search completes within $\sim$1~second for a single GEMV, 2--3~seconds for a single GEMM, and 5--10~seconds for LLM workloads on a 16-core consumer-grade CPU. 
The search space includes 1,548 candidate mappings for a single GEMM and 192 for GEMV, which has a fixed dimension of 1 require much less tiling and mapping options.The search is very fast as: 
(1) each evaluation relies on an analytical model and is completed within microseconds, and 
(2) LLM workloads use consistent GEMM and GEMV shapes across layers, allowing optimal mappings to be reused. Moreover, the search overhead can be pre-paid or amortized. For LLM inference workloads, matrix dimensions vary only with input token length, but because mapping is data-agnostic, mappings for different token lengths can be precomputed or cached at runtime, effectively eliminating repeated search cost.

\textbf{Integration of Mapping Framework} Our mapping framework can be integrated with Front-end / high-level IR such as PyTorch~\cite{pytorch}, TensorFlow~\cite{tensorflow}, MLIR~\cite{mlir}, Halide~\cite{halide}  TVM Relay~\cite{tvm} etc. by annotating PIM-eligible ops and precision.  Our proposed framework can act as a mapping pass and codegen backend for lowering IR to PIM instructions, and insert data layout transformations where necessary.

%% file: sec/08_related_work.tex
\section{Related Work}
\label{sec:related_work}
To the best of our knowledge, \sysname ~is the first bit-serial PIM architecture that enables bit-level data reuse and supports a mapping framework that explores the entire mapping space. Here, we highlight our contributions by comparing with state-of-the-art PIM works shown in table.\ref{tab:arch_comparison}. 

\textbf{Processing Using DRAM} Prior work~\cite{simdram,mimdram,proteus,ambit, drisa, rowclone, dracc, elp2im, computeDRAM, pluto, drange, scope, mimdram, chopper, lacc, flexidram} extend in-DRAM mechanisms like row-copy and bulk bitwise operations to implement logic operations using standard DRAM commands~\cite{ambit,bulk_bitwise,rowclone}. These processing-using-DRAM (PUD) systems do not require the integration of additional processing elements (PEs) into DRAM and, in some cases, employ bit-serial schemes to support dynamic bit precision. However, these systems typically offer limited or no data reuse across the system and rely on manually tuned or heuristic-based workload mapping. As a result, existing work~\cite{proteus} shows poor performance on GEMM workloads (2mm, 3mm, gmm from PolyBench\cite{polybench}) compared to GPUs.

\textbf{Digital PIM} Processing-in-Memory systems with digital PEs are referred to as digital PIM. SRAM-based systems such as NeuralCache~\cite{neural_cache,pimsab} and PIMSAB incorporate bit-serial PEs at SRAM arrays to enable computation~\cite{neural_cache,pimsab}, but still intrinsically limited by DRAM bandwidth. Even DRAM-based systems~\cite{newtown_AiM,AttenPIM, neupims, BlockPIM, DEAR-PIM} add near Bank/Subarray PEs, the parallelism are bounded by the DRAM column decoder. Furthermore, the PEs support only fixed precision, limiting the adoption of mixed-precision optimizations. Other systems \cite{SeIM, SAL-PIM,PIMPAL} improve the DRAM search/lookup capability by adding simple logic across memory hierarchies. However, such systems usually support fixed workloads, limiting the scope of broader applicability. 
\sysname adds light-weight bit-serial PEs close to arrays, explores sub-array-level internal bandwidth, and supports broad workloads and flexible precisions.
% hardly {\color{red}capture} data locality within the memory, leading to significant data movement overhead as well as excessive accesses.\textcolor{red}{The last sentence doesn't feel very correct. Need to find other aspects to place our work among those. Response: I just changed explore to capture. Is it better now?}

\textbf{Mapping} Prior works ~\cite{tvm, timeloop, sparseloop, mind_mappings, ruby, maeri, sod2,  interstellar, cosa, dnn_lp, gamma, digamma, wang2021searchoptimalsystolicarrays, expdse, cocco, triton, tensorcomprehensions, marvel, tenet} present loop-analysis-based mapping techniques or frameworks for ASIC accelerators. However, they do not generalize well to in-DRAM processing architectures due to hierarchical structure and sub-array-aware data layout.
Several works\cite{pimdl, ares, optipim} propose frameworks to abstract and search the PIM mapping space for loop-based workloads. PIM-DL and ARES use genetic algorihms and OptiPIM use Integer Linear Programming to search over the mapping space. Comparing to those works, \sysname's mapping abstraction limit its scope to the specific mapping space of GEMM, focusing on mapping GEMM dimensions to DRAM's hierarchical architecture. Combined with an analytical performance model, \sysname~ enables searching over the entire mapping space and find the optimal mapping.

% \textcolor{red}{more PIM mapping to add}

% Define shorter commands for cleaner code
\newcommand{\cmark}{\ding{51}}  % ✓ 
\newcommand{\xmark}{\ding{55}}  % ✗
\begin{table}[t]
\centering
\caption{Comparison of architecture design and mapping methodologies. Assume n bit operands.}
\label{tab:arch_comparison}
\renewcommand{\arraystretch}{1.1}
\setlength{\tabcolsep}{2.5pt} % slightly tighter columns

\resizebox{\columnwidth}{!}{
\begin{tabular}{lccccc>{\raggedright\arraybackslash}p{2.3cm}}
\toprule
\textbf{Prior Works} &
\makecell{\textbf{Compute}\\\textbf{Scheme}} &
% \makecell{\textbf{Row accesses }\\\textbf{per N-bit}\\\textbf{multiplication}} &

\makecell{\textbf{Row ACTs} \\\textbf{of n-bit Mult}}&
\makecell{\textbf{Broadcast}\\\textbf{Hardware}} &
\makecell{\textbf{Reduction}\\\textbf{Hardware}} &
\makecell{\textbf{Mapping}\\\textbf{Methodology}} \\
\midrule
Neural Cache & SRAM, bit-serial & -- & \cmark & \xmark & Manual \\
PIMSAB       & SRAM, bit-serial & -- & \cmark & \cmark & Heuristics \\
Newton       & DRAM, bit-parallel & $O(n^2)$ & \cmark & \cmark & Manual \\
SIMDRAM      & DRAM, bit-serial & $O(n^2)$ & \xmark & \xmark & Manual \\
MIMDRAM      & DRAM, bit-serial & $O(n^2)$ & \xmark & \xmark & Heuristics \\
Proteus      & DRAM, bit-serial & $O(n^2)$ & \xmark & \xmark & Manual \\

\sysname(Ours)         & DRAM, bit-serial & $O(n)$ & \cmark & \cmark & Exhaustive Search \\
\bottomrule
\end{tabular}
}
\end{table}

%% file: sec/09_conclusion.tex
%\vspace{5mm}
\section{Conclusion}
\label{sec:conclusion}
\vspace{3mm}
We introduce \sysname, a DRAM based Processing-In-Memory system with bit level reuse and broadcast capabilities. Prior processing in memory architectures struggled with matrix multiplications, while  \sysname~ captures locality and significantly accelerates matrix operations and large language models. To achieve full bit level reuse, \sysname ~adds locality buffer to store some bits of operands during integer multiplication.  To reduce data reorganization overhead between LLM kernels, \sysname ~adds broadcasting units. To achieve efficient reduction, a common operation in LLM kernels, \sysname ~adds reduction units at each subarray to accelerate column-wise reduction. To efficiently map all matmul kernels of LLM, \sysname ~proposes a mapping framework to automatically search for the optimal mapping strategy for each matmul kernel. We evaluate \sysname ~against the H100 GPU and SOTA DRAM PIM  systems, and demonstrate improvements in both performance and performance per area.
%\sysname~ delivers an average of 90.1x and 15.6x performance improvement over GPUs on two end-to-end inference senarios and achieves an average of $46\times$ and $233\times$ performance/$mm^2$ improvement over GPUs and SOTA PIMs respectively, with around 4\% chip area overhead. 
\noindent\sysname ~delivers an average of 90.1x and 15.6x performance improvement over GPUs on two end-to-end inference senarios with GPT3 and Llama3 workloads. Likewise, \sysname~ achieves an average of $46\times$ over GPUs for performance/$mm^2$ and  improvement (geomean) of more than $200\times$  over SOTA PIMs with only approximately 4\% chip area overhead.
% We open-source our simulator source code at XXX.
% We introduce \sysname, a DRAM based Processing-In-Memory system with full bit level reuse and accelerates both prefill and decode of large language models. To achieve full bit level reuse, \sysname adds locality buffer to store some bits of operands during integer multiplication.  To reduce data reorganization overhead between LLM kernels, \sysname adds broadcasting units. To achieve efficient reduction, a common operation in LLM kernels, \sysname adds reduction units at each subarray to accelerate column-wise reduction. To efficiently map all matmul kernels of LLM, \sysname propose a mapping framework to automatically search for the optimal mapping strategy for each matmul kernel. We evaluate \sysname against H100 and Proteus systems, and demonstrate improvments in both performance and performance per area. We open-source our simulator source code at XXX.